\documentclass[pdflatex,sn-nature,Numbered]{sn-jnl}

\usepackage{graphicx}%
\usepackage{multirow}%
\usepackage{amsmath,amssymb,amsfonts}%
\usepackage{amsthm}%
\usepackage{mathrsfs}%
\usepackage[title]{appendix}%
\usepackage{xcolor}%
\usepackage{xspace}%
\usepackage{textcomp}%
\usepackage{manyfoot}%
\usepackage{booktabs}%
\usepackage{algorithm}%
\usepackage{algorithmicx}%
\usepackage{algpseudocode}%
\usepackage{listings}%
\usepackage{subcaption}
\usepackage{anyfontsize}
\usepackage{natbib}

\newcommand\aap{A\&A}                
\newcommand\aj{AJ}                   
\newcommand\apj{ApJ}                 
\newcommand\apjl{ApJ}                
\newcommand\apjs{ApJS}               
\newcommand\araa{ARA\&A}             
\newcommand\mnras{MNRAS}             
\newcommand\nat{Nature}              
\newcommand\pasp{PASP}               

\newcommand{\kms}{\ensuremath{\mathrm{km\,s^{-1}}}\xspace}

\begin{document}

\title[Jekyll, Hyde and co.]{Accelerated quenching and chemical enhancement of massive galaxies in a $z\sim4$ gas-rich halo}



\author*[1]{\fnm{Pablo G.} \sur{P\'erez-Gonz\'alez}}\email{pgperez@cab.inta-csic.es}
\equalcont{These authors contributed equally to this work.}

\author[2,3]{\fnm{Francesco} \sur{D'Eugenio}}\equalcont{These authors contributed equally to this work.}

\author[1]{\fnm{Bruno} \sur{Rodr\'{\i}guez del Pino}}

\author[1]{\fnm{Michele} \sur{Perna}}

\author[2,3,4]{\fnm{Hannah} \sur{\"Ubler}}

\author[2,3]{\fnm{Roberto} \sur{Maiolino}}

\author[1]{\fnm{Santiago} \sur{Arribas}}

\author[5]{\fnm{Giovanni} \sur{Cresci}}

\author[5,1]{\fnm{Isabella} \sur{Lamperti}}


\author[6]{\fnm{Andrew~J.} \sur{Bunker}}

\author[7]{\fnm{Stefano} \sur{Carniani}}

\author[8]{\fnm{Stephane} \sur{Charlot}}

\author[9]{\fnm{Chris J.} \sur{Willott}}


\author[10]{\fnm{Torsten} \sur{B\"{o}ker}}


\author[7]{\fnm{Eleonora} \sur{Parlanti}}

\author[2,3]{\fnm{Jan} \sur{Scholtz}}

\author[7]{\fnm{Giacomo} \sur{Venturi}}


\author[11]{\fnm{Guillermo} \sur{Barro}}

\author[1]{\fnm{Luca} \sur{Costantin}}

\author[12,13]{\fnm{Ignacio} \sur{Mart\'{\i}n-Navarro}}

\author[14]{\fnm{James S.} \sur{Dunlop}}

\author[15]{\fnm{Daniel} \sur{Magee}}

\affil[1]{Centro de Astrobiolog\'{\i}a (CAB), CSIC-INTA, Ctra. de Ajalvir km 4, Torrej\'on de Ardoz, E-28850, Madrid, Spain}

\affil[2]{Kavli Institute for Cosmology, University of Cambridge, Madingley Road, Cambridge, CB3 OHA, UK}

\affil[3]{Cavendish Laboratory - Astrophysics Group, University of Cambridge, 19 JJ Thomson Avenue, Cambridge, CB3 OHE, UK}

\affil[4]{Max-Planck-Institut f\"ur extraterrestrische Physik, Gießenbachstraße 1, 85748 Garching, Germany}

\affil[5]{INAF - Osservatorio Astrofisco di Arcetri, Largo E. Fermi 5, 50127, Firenze, Italy}

\affil[6]{Department of Physics, University of Oxford, Denys Wilkinson Building, Keble Road, Oxford OX1 3RH, UK}

\affil[7]{Scuola Normale Superiore, Piazza dei Cavalieri 7, I-56126 Pisa, Italy}

\affil[8]{Sorbonne Universit\'e, CNRS, UMR 7095, Institut d'Astrophysique de Paris, 98 bis bd Arago, 75014 Paris, France}

\affil[9]{NRC Herzberg, 5071 West Saanich Rd, Victoria, BC V9E 2E7, Canada}

\affil[10]{European Space Agency, c/o STScI, 3700 San Martin Drive, Baltimore, MD 21218, USA}

\affil[11]{Department of Physics, University of the Pacific, Stockton, CA 90340 USA}

\affil[12]{Instituto de Astrof\'{\i}sica de Canarias, C/ V\'{\i}a L\'actea s/n, E38205- La Laguna, Tenerife, Spain}

\affil[13]{Departamento de Astrof\'{\i}sica, Universidad de La Laguna, E-38200 La Laguna, Tenerife, Spain}

\affil[14]{Institute for Astronomy, University of Edinburgh, Royal Observatory, Edinburgh EH9 3HJ, UK}

\affil[15]{UCO/Lick Observatory, University of California, Santa Cruz, CA 95064, USA}


\abstract{\textbf{Stars in galaxies form when baryons radiatively cool down and fall into gravitational wells whose mass is dominated by dark matter. Eventually, star formation quenches as gas is depleted and/or perturbed by feedback processes, no longer being able to collapse and condense. We report the first spatially resolved spectroscopic observations, using the JWST/NIRSpec IFU, of a massive, completely quiescent galaxy (Jekyll) and its neighborhood at $z=3.714$, when the Universe age was 10\% of today's. Jekyll resides in a massive dark matter halo (with mass M$_\mathrm{DM}>10^{12}$~M$_\odot$) and forms a galaxy pair with Hyde, which shows very intense dust-enshrouded star formation (star formation rate $\sim300$~M$_\odot$~yr$^{-1}$). We find large amounts of kinematically perturbed ionized and neutral gas in the circumgalactic medium around the pair. Despite this large gas reservoir, Jekyll, which formed $10^{11}$~M$_\odot$ in stars and chemically enriched early (first billion years of the Universe) and quickly (200--300~Myr), has remained quiescent for over 500~Myr. The properties of the gas found around the two galaxies are consistent with intense, AGN-induced photoionization, or intense shocks. However, with the current data no obscured or unobscured AGN is detected in the central galaxy (Jekyll) nor in the very active  and dust rich star-forming galaxy (Hyde).}}



\keywords{massive galaxies, galaxy groups, cosmological paradigm, JWST, Universe enrichment, early Universe}

\maketitle

As a basic test of the $\Lambda$CDM paradigm, astronomers have been searching for massive galaxies at high redshift for more than two decades. There is particular interest in finding and confirming the nature of those that are not forming stars anymore, i.e., they are quiescent, which implies that they formed a significant fraction of their large stellar content at very early cosmic epochs \cite{2008A&A...482...21C,2009ApJ...697.1290B,2009ApJ...705L..71K,2016MNRAS.457.3743D,2017Natur.544...71G,2018MNRAS.480.4379C,2024Natur.628..277G}. 


The quick and early appearance of massive  galaxies must be linked to an enhanced infall of baryons into potential wells (whose mass is dominated by dark matter and are known as dark matter halos) and/or a more efficient transformation of gas into stars \cite{2023MNRAS.523.3201D,2023MNRAS.522.3986F}, which has been confirmed by the discovery with JWST of many bright galaxies at very high redshifts \cite{2022ApJ...940L..55F,2023MNRAS.518.6011D,2023ApJ...951L...1P,2023arXiv230406658H}. The quenching of the star formation in those massive galaxies must be linked, in contrast, with some negative feedback mechanisms that cut the gas supply or inhibit its transformation into stars \cite{2005MNRAS.361..776S,2006MNRAS.365...11C,2012ARA&A..50..455F,2012MNRAS.421.3464P,2014MNRAS.445..581H}. The search and detailed analysis of the expeditiously forming and also early-quenched galaxies is, therefore, of utmost importance to understand how star formation behaves in the Universe by studying the positive feedback or feedback-free early stellar mass assembly and the negative feedback mechanisms that halt star formation and maintain it switched off for long times.




One notable example of this effort to study massive quiescent systems is the galaxy ZF-COSMOS-20115 \citep{2017Natur.544...71G}, later called `Jekyll' in \cite{2018A&A...611A..22S}. Jekyll qualified as the highest redshift ($z=3.717$, corresponding to Universe age of 1.6~Gyr, 11\% of its current age) massive quiescent galaxy at the time of its discovery. Although some other quiescent massive galaxies at higher redshifts have been reported, Jekyll remains the most distant to date without any sign of recent star formation. Indeed, its specific star formation is $0.05\pm0.03$~Gyr$^{-1}$, i.e., it would need 13 times the age of the Universe at $z=3.7$ to form their stellar content at the current rate, implying a much higher star formation activity in the past. Jekyll does not show any nuclear activity either, as we will show in this paper, in contrast with, e.g., the examples presented in \cite{2023arXiv230111413C} or \cite{2024arXiv240405683D}.

The apparently quiescent nature of Jekyll was challenged by the detection of dust continuum emission with SCUBA-2. But those observations as well as the detection of [CII] ionized gas with ALMA were demonstrated to come from a separate nearby source at the same redshift \citep{2017ApJ...844L..10S,2018A&A...611A..22S}. The intense emission from [CII] that was reported is classically linked to photo-dissociation regions around recent star-formation sites or regions around AGN (see, e.g., \cite{2010ApJ...724..957S,2018ApJ...854...97D,2021A&A...654A..37D}). Further analysis finally arrived to the conclusion that there were two galaxies in this system: Jekyll, the massive quiescent galaxy, and a dusty starburst galaxy, dubbed Hyde, separated by 3~kpc (projected distance). 


In this paper, we present integral-field spectroscopy (IFS) of the Jekyll \& Hyde system obtained with the JWST/NIRSpec instrument \cite{2022A&A...661A..82B}, which allows us to discover new components of this group of massive galaxies at $z\sim4$ and help to understand their properties in two dimensions, and the story of its evolution \cite[see also][]{2023arXiv231003067P}.



This system (hereafter: the {\it Tusitala} Group) is used here to demonstrate three main results: (1) Passively evolving, metal-enriched,  massive galaxies formed expeditiously and exist earlier than what state-of-the-art simulations predict. (2) Even though the immediate neighborhood of those massive galaxies is gas rich, they  manage to remain quiescent for hundreds of million years. (3) Massive galaxies (in our case, a pair of them) can be surrounded by significant amounts of ionized gas presenting kinematic structures and emission-line ratios compatible with tidal disruption, AGN photoionization and shocks.


The novel and unique JWST IFS dataset used in this paper includes prism observations with low spectral resolution ($R\sim100$) and wide spectral range ($\lambda=0.6-5.2$~$\mu$m), and grating observations with high spectral resolution ($R\sim2700$), covering the H$\alpha$ and H$\beta$+[OIII] spectral ranges (from 1.6 to 3.1~$\mu$m). The spatial resolution of the NIRSpec IFS data ranges from FWHM $\simeq$ 0.1'' to 0.18'', allowing us to probe this high-redshift system with spatially resolved spectroscopy at sub-galactic, kiloparsec physical scales within a region of $15\times15$~kpc$^2$ (in contrast with other studies based on slit spectroscopy, lacking the 2D coverage achieved with IFS).

The JWST/NIRSpec observations reveal a very complex structure in this group of high-redshift galaxies. This is shown in Fig.~1. Apart from the Jekyll and Hyde galaxies located in the center of the image and whose accurate redshift determinations, kinematics analysis, and dust content are presented in Figure~2 and Extended Data Figures~1, 2, 3, 4, and 5, several more components are detected in the JWST/NIRSpec data extending over a 225~kpc$^2$ area. The NIRSpec observations reveal emission lines (the brightest being [OIII] and H$\alpha$, the latter depicted in the figure in green) arising from an extended region (5~kpc diameter) located to the Northeast of the most massive galaxy (Jekyll). We dub this region Eastfield. A very elongated region (6~kpc long, 3~kpc wide, extending along the NW to SW direction) with emission lines is located to the West, dubbed Mr~West. Emission line ratio properties are presented in Extended Data Figure~6 and Methods section~3.2.



In contrast, imaging data provided by the JWST/NIRCam instrument barely detect the brightest object, Jekyll, at observed wavelengths bluer than $\sim$1.5~$\mu$m (as an example, we show the F115W band in the top-right panel of Figure~1). Jekyll becomes very bright at observed wavelengths longer than 2~$\mu$m, while Hyde starts to be detected clearly at the longest wavelengths ($\gtrsim3$~$\mu$m). Both galaxies are very red for different reasons (stellar age and dust content; see Extended Data Figures~7, 8, and 9, and discussion about their stellar content in Methods Section~4). Eastfield and Mr~West only appear as very diffuse, ultra-faint, extended emission in the broad-band imaging data obtained with the NIRCam instrument, especially in the bands enclosing bright emission lines, indicating that they are gas-rich systems with relatively small stellar content.


The kinematic analysis performed with the high spectral resolution JWST/NIRSpec data, as well as the stellar population synthesis modeling carried out with the low spectral resolution observations (also the $R=2700$ data for the core of Jekyll), are used to study the physical properties and nature of the four components of the {\it Tusitala} group mentioned above (and the one presented in the next paragraph). Fig.~2 shows the kinematics of the system, Fig.~3 shows the star formation history (SFH) of the two main galaxies (Jekyll and Hyde), and Fig.~4 shows maps of the main physical properties (stellar mass, mass-weighted age, specific star formation rate, and dust attenuation). 

Apart from emitting material, a fifth component is revealed in our analysis for the first time: a dark cloud in the foreground of Jekyll, whose composition includes neutral sodium and is only detectable through the absorption in the spectrum of the massive quiescent galaxy, as shown in Fig.~5. We call this component Dr~Sodium. Its velocity is blueshifted by 200~\kms with respect to Jekyll and has a mass of $\sim10^{8.5}$~M$_\odot$. Dr~Sodium is observed across the entire region where back-illumination from Jekyll is detected, which suggests its full extent may be larger. If interpreted as an outflow from Jekyll, it would give a mass-outflow rate $>250$~M$_\odot$~yr$^{-1}$.



Our analysis of the {\it Tusitala} Group reveals a complex history and structure history, which we summarize in the cartoon presented in Fig.~6. Its most massive member, Jekyll, followed a closed-box scenario for its formation \cite{2002ApJ...581.1019G}. The main star formation episode occurred around $z=5-6$, around 1~Gyr after the Big Bang, managing to form  $10^{11}$~M$_\odot$ in less than 300~Myr (see Figure~3 and Extended Data Figure~10). The age gradient indicates an inside-out formation \citep{2018ApJ...859...56T}; the significant metal enrichment, expressed in the nearly solar metallicity and the high $\alpha$-element enhancement at its core (see Methods section 4.3, and Extended Data Figures~8 and 9) reveals a quick and efficient (relatively free of negative feedback) transformation of the initially collapsed gas into stars with little replenishment (and metal dilution), and an eventual stop due to fuel exhaustion \citep{2020MNRAS.491.5406T}. After consuming all its gas (and maybe preventing more gas to infall due to AGN feedback), Jekyll became quiescent.



The most interesting result of our study resides on how Jekyll, after its early and rapid formation, has stayed dead for more than half a Gyr. The {\it Tusitala} Group reveals that the circum- and intergalactic medium and environment around Jekyll is not gas free. We detect gas-rich nebulae around Jekyll, i.e., the Eastfield and Mr~West components. These two nebulae are aligned with another massive galaxy, Hyde. 
Hyde also presents significant metal enrichment, this time expressed in its high dust content (see Extended Data Figure~1 and Methods Section~4). The level of star formation in Hyde in the last 100~Myr is similar to the peak activity of Jekyll (around 300~M$_\odot$~yr$^{-1}$).


The dusty star-forming galaxy Hyde is experiencing two effects that most probably will end up in its own quenching, which our estimations of the SFH reveals that has already started. Remarkably, one of these effects is unique for this galaxy, since it is linked to the presence of Jekyll.


The first effect is observed in the properties of Eastfield, most remarkably, but also in Mr~West. These are gas-rich extended and diffuse components of the group. We calculate a mass of ionized gas in these systems around $\sim10^{9}$~M$_\odot$ (see Methods and \cite{2023arXiv231003067P}). Their rest-frame optical spectra are dominated by emission lines from hydrogen, oxygen, and nitrogen, arising from ionized gas, revealing that it has been enriched by outflows from the central galaxies, or the gas has been extracted from their interstellar medium, or there is {\it in situ} star formation in these neighbor systems. The line ratios of this gas are consistent with excitation by AGN and shocks \cite{1981PASP...93....5B,2016ApJS..224...38A}. The faint continuum detected in the spectroscopic and photometric data provided by the JWST NIRSpec and NIRCam instruments implies a very low content of stars. In fact, assuming the continuum flux is due to stellar emission, we obtain stellar masses around 10$^9$~M$_\odot$ for each component, Eastfield and Mr~West, with the brightest emission-line knots presenting stellar masses below $10^{8}$~M$_\odot$. The measured fluxes could also be entirely due to nebular continuum emission and, consequently, those stellar mass estimates should be regarded as upper limits. In addition, the velocity dispersion in several regions of Eastfield and Mr~West is high (200--300~\kms). All these properties indicate that these kinematically complex regions are nebulae most probably perturbed (photoionized and maybe shocked) by the activity in Hyde and/or an {\it in situ} AGN (as proposed by \cite{2023arXiv231003067P}), given that Jekyll has been quiescent for $\sim0.5$~Gyr. 





The second effect that can contribute to quench Hyde and that is unique to this galaxy compared to what Jekyll suffered when it halted its star formation, is linked to tidal effects induced by the gravitational interactions of Hyde with the more massive galaxy Jekyll. This effect is seen especially in one of the gas-rich components of the {\it Tusitala} Group, Mr~West. The elongated morphology connecting it with Hyde and the large velocity range observed for this system are completely consistent with a tidal tail, which would be removing gas from Hyde as it approached Jekyll in the recent past \cite{1972ApJ...178..623T,2013LNP...861..327D}.





The overall conclusion is that after a major starburst in the central galaxy of a massive halo, Jekyll is remaining quiescent despite its still gas-rich environment. In contrast, another very close massive galaxy companion, Hyde, is still consuming large amounts of gas itself. The large gas reservoir around these two galaxies detected in our JWST/NIRSpec IFU data presents extreme ionization properties, consistent with AGN-driven energy injection (photoionization or shocks). However, no nuclear activity has been detected anywhere in the group, but the presence of an obscured AGN in Hyde,  or a faded one which is echoing on Eastfield and Mr~West (similar to some systems in the nearby Universe, \cite{2009MNRAS.399..129L,2012AJ....144...66K}), cannot be ruled out or confirmed with the current data. 




\begin{figure}
\centering
\begin{minipage}{0.7\textwidth}
\includegraphics[trim=3.1cm 2.7cm 2.5cm 3.0cm,clip,width=1.0\textwidth]{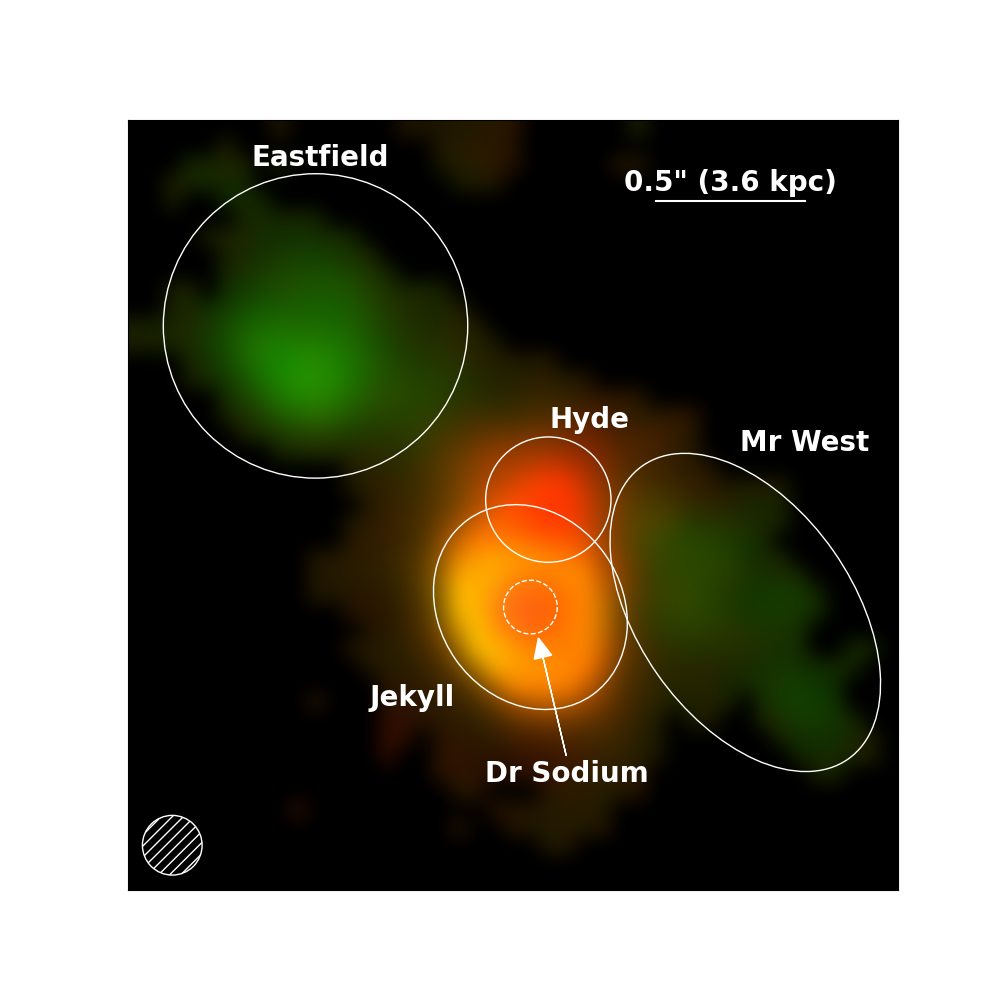}\\
\end{minipage}
\begin{minipage}{0.25\textwidth}
\vspace{-0.4cm}
\includegraphics[trim=3.1cm 2.7cm 2.5cm 3.0cm,clip,width=0.9\textwidth]{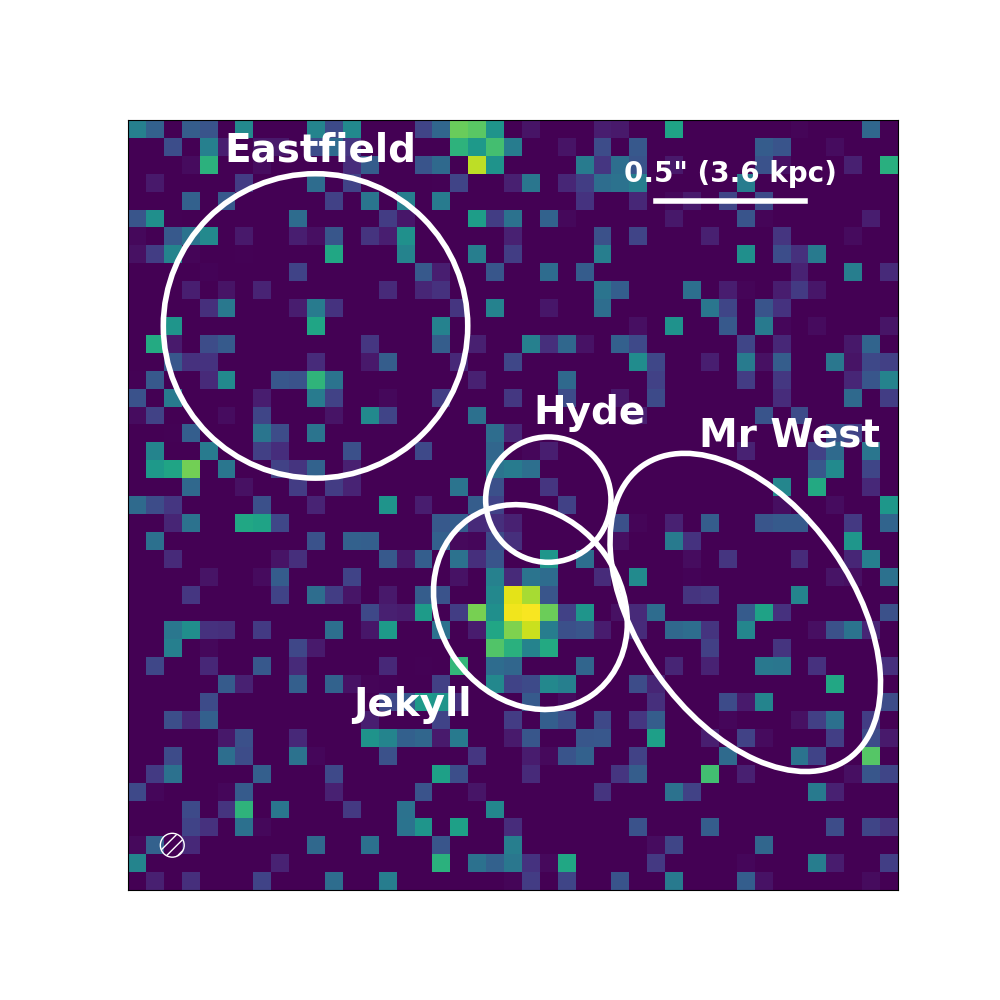}
\includegraphics[trim=3.1cm 2.7cm 2.5cm 3.0cm,clip,width=0.9\textwidth]{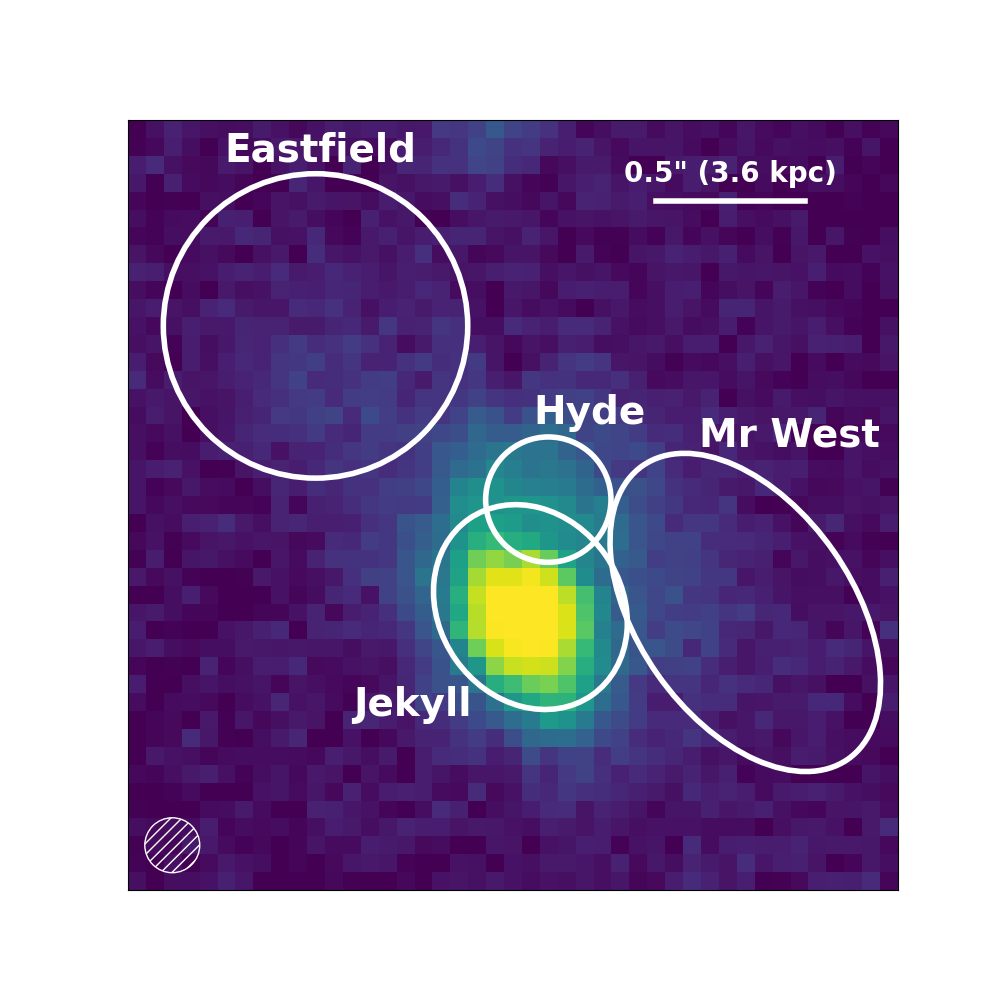}
\includegraphics[trim=3.1cm 2.7cm 2.5cm 3.0cm,clip,width=0.9\textwidth]{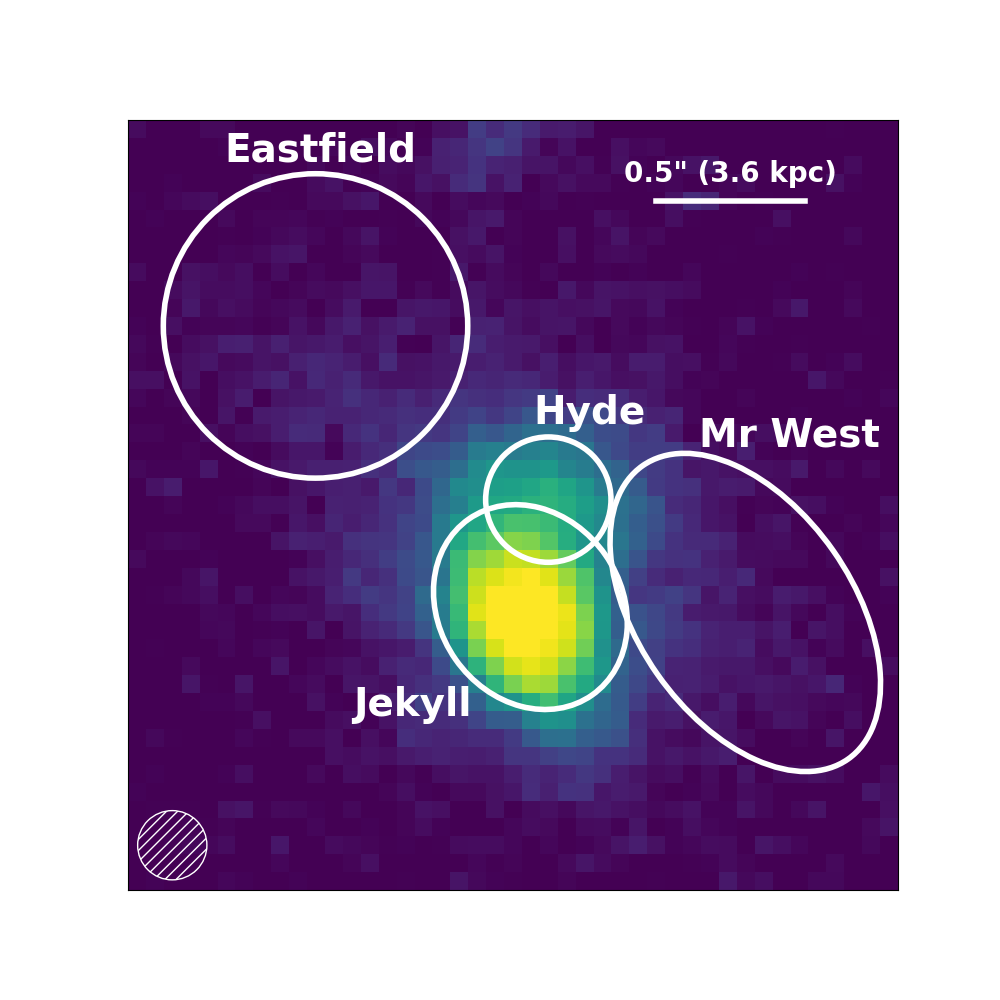}
\end{minipage}
\caption{\label{fig:rgb}{\bf RGB images of the {\it Tusitala} system, a galaxy pair in a gas-rich halo at $z=3.7$.} The main panel shows an RGB image constructed with the JWST/NIRSpec dataset. The system comprises two main massive galaxies called Jekyll and Hyde, and three additional gas-rich components. The image is constructed with F277W (red), F200W (blue) and H$\alpha$ (green) synthetic images built from the JWST/NIRSpec prism (spectral resolution $R=100$) integral field spectroscopic observations. The new JWST data reveal an evolved group of galaxies formed 1.6~Gyr after the Big Bang, whose distinct colors in this RGB image demonstrate their very different nature and/or evolutionary state. The galaxy system components, marked with ellipses and circles in the figure, are: (1) a massive quiescent galaxy, presenting red to orange colors, dubbed Jekyll; (2) a dusty massive starburst to the North separated at least (i.e., in projection) 3~kpc from Jekyll, presenting also a red color, dubbed Hyde; and (3) and (4) two ionized nebulae presenting a green color as their optical spectra are dominated by emission lines, dubbed Eastfield and Mr~West, separated 6-10~kpc from the central Jekyll galaxy. Another component, (5), is also present in this system, which is only revealed by absorption in the spectrum of Jekyll (being located in its foreground) and we call it Dr~Sodium (shown with a circle in the core of Jekyll, for clarity). The right panels show postage stamps in NIRCam filters covering the short and long wavelength spectral ranges. From top to bottom, the bands are: F115W, F277W, and F356W (Dr~Sodium is not shown in these stamps for clarity). All the figures show the angular resolution of the data with a filled circle (varying across the covered wavelength range, and homogenized to the reddest wavelengths in the case of the NIRSpec data). We also depict the angular and physical scale in the top right corner, as well as ellipses/circles enclosing the five components of the group.}
\end{figure}

\begin{figure}
\centering
\includegraphics[trim=12.6cm 2.2cm 9.0cm 5.0cm,clip,width=1.0\textwidth]{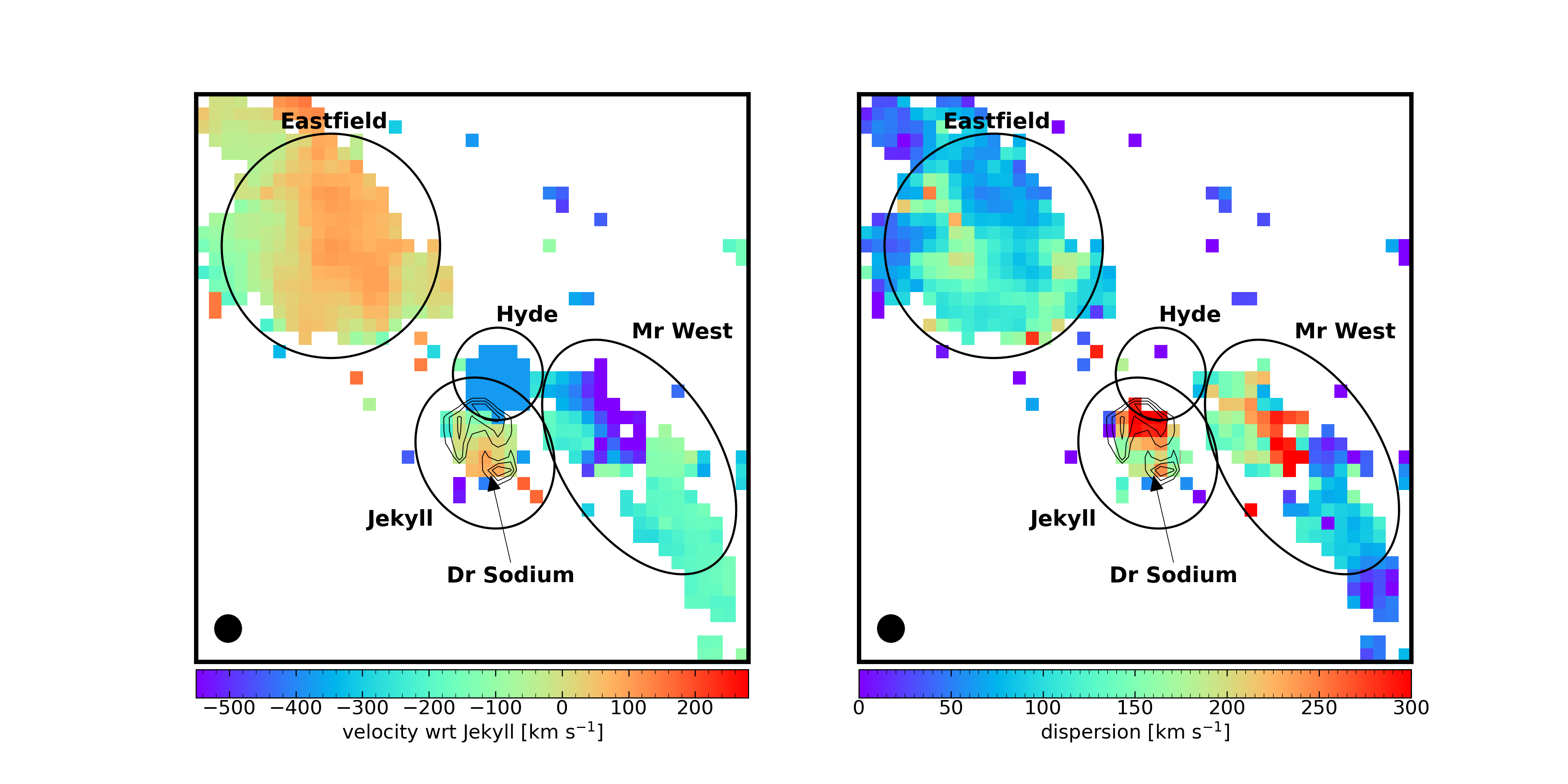}
\caption{\label{fig:kinematics}{\bf Kinematics of the {\it Tusitala} group.} The plot depicts all members of the group, whose 2 main components are the Jekyll and Hyde massive galaxies. On the left, velocities with respect to Jekyll (reference redshift $z=3.7135$). On the right, we show the dispersion map (no robust measurements for Hyde can be performed). For Eastfield and Mr~West, the kinematics measurements come from nebular emission lines ([OIII] and H$\alpha$) measured in the $R=2700$ cube. For Jekyll, the kinematics come from stellar absorption lines detected in the same dataset. For Hyde, the velocity measurement corresponds to the H$\alpha$ emission line detected in the $R=100$ spectrum with $S/N>8$ and in the $R=2700$ with $S/N\sim4$. The contours show the equivalent width of the NaD absorption in the Jekyll $R=2700$ spectra (3 contours showing 5, 7, and 9~\AA), indicating the position of Dr. Sodium. The different components of the system are enclosed in the same elliptical and circular regions shown in Figure~1, the spatial resolution (FWHM of the PSF) is also shown in the bottom-left corner.}
\end{figure}

\begin{figure}
\centering
\includegraphics[width=1.0\textwidth]{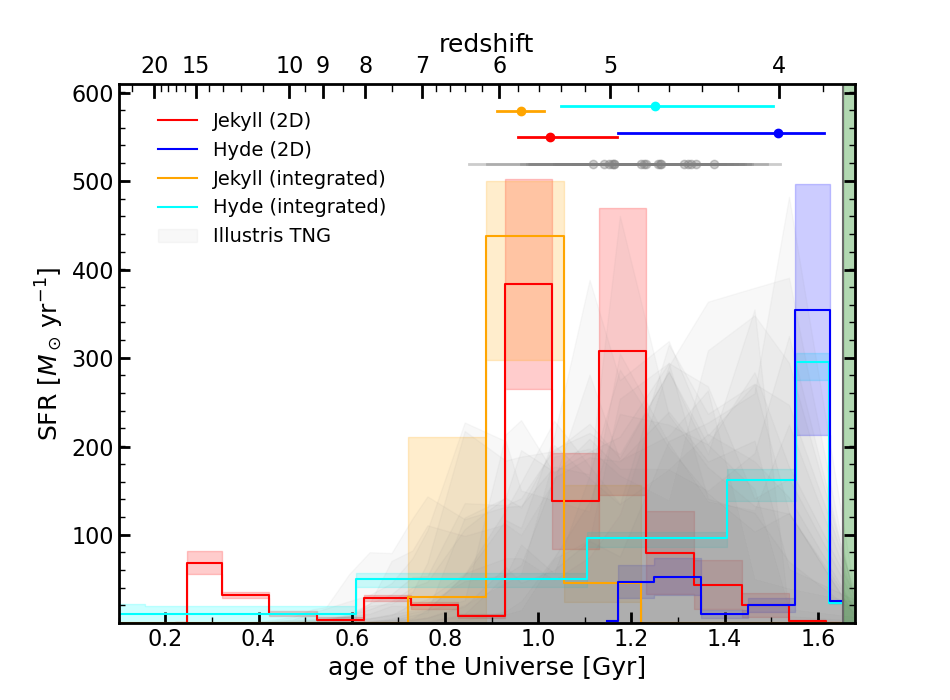}
\caption{\label{fig:sfh}{\bf Star formation history of the Jekyll and Hyde galaxies, the two massive galaxies in the {\it Tusitala} group.} The calculations are based on the 2-dimensional analysis of the stellar populations carried out with SED-fitting to the $R=100$ data (in red and blue for Jekyll and Hyde, respectively), as well as a non-parametric approach to the analysis of their integrated SED (in orange and cyan). The green shaded region marks the epoch after the system redshift (assumed to be that of Jekyll, $z=3.7135$). Gray areas show the star-formation histories of massive ($\mathrm{M_\star}>10^{11}$~M$_\odot$) galaxies in the Illustris TNG simulations (considering all the most massive dark matter halos in the simulations for all available resolutions, namely TNG-50, TNG-100, and TNG-300). Only galaxies with star-formation rate lower than 50~M$_\odot$~yr$^{-1}$ are shown (around 8\% of the full sample of massive galaxies in TNG), the minimum being 20~M$_\odot$~yr$^{-1}$, corresponding to a specific star formation rate of 0.1~Gyr$^{-1}$, similar to our measurements for Jekyll. The points and segments at the top show the mass-weighted ages, and times around it where 50\% of the total stellar mass of each galaxy was formed.}
\end{figure}

\begin{figure}
\centering
\includegraphics[trim=20.3cm 22.2cm 18.0cm 8.0cm,clip,width=0.9\textwidth]{images/j_and_h_and_co_spsmaps_v3.4.pdf}
\caption{\label{fig:spsmaps}{\bf Maps of the physical properties of the Jekyll and Hyde galaxy system.} They have been obtained from the SED-fitting analysis of the $R=100$ spectro-photometric data. Clockwise from the top left, we show the stellar mass, specific star formation rate, dust attenuation, and mass-weighted age. Ellipses and circles are the same shown in Figures~1 and 2, the spatial resolution (FWHM of the PSF) is also shown in the bottom-left corner.}
\end{figure}

\begin{figure}
\centering
\includegraphics[trim=0.0cm 0.0cm 0.0cm 0.0cm,width=0.8\textwidth]{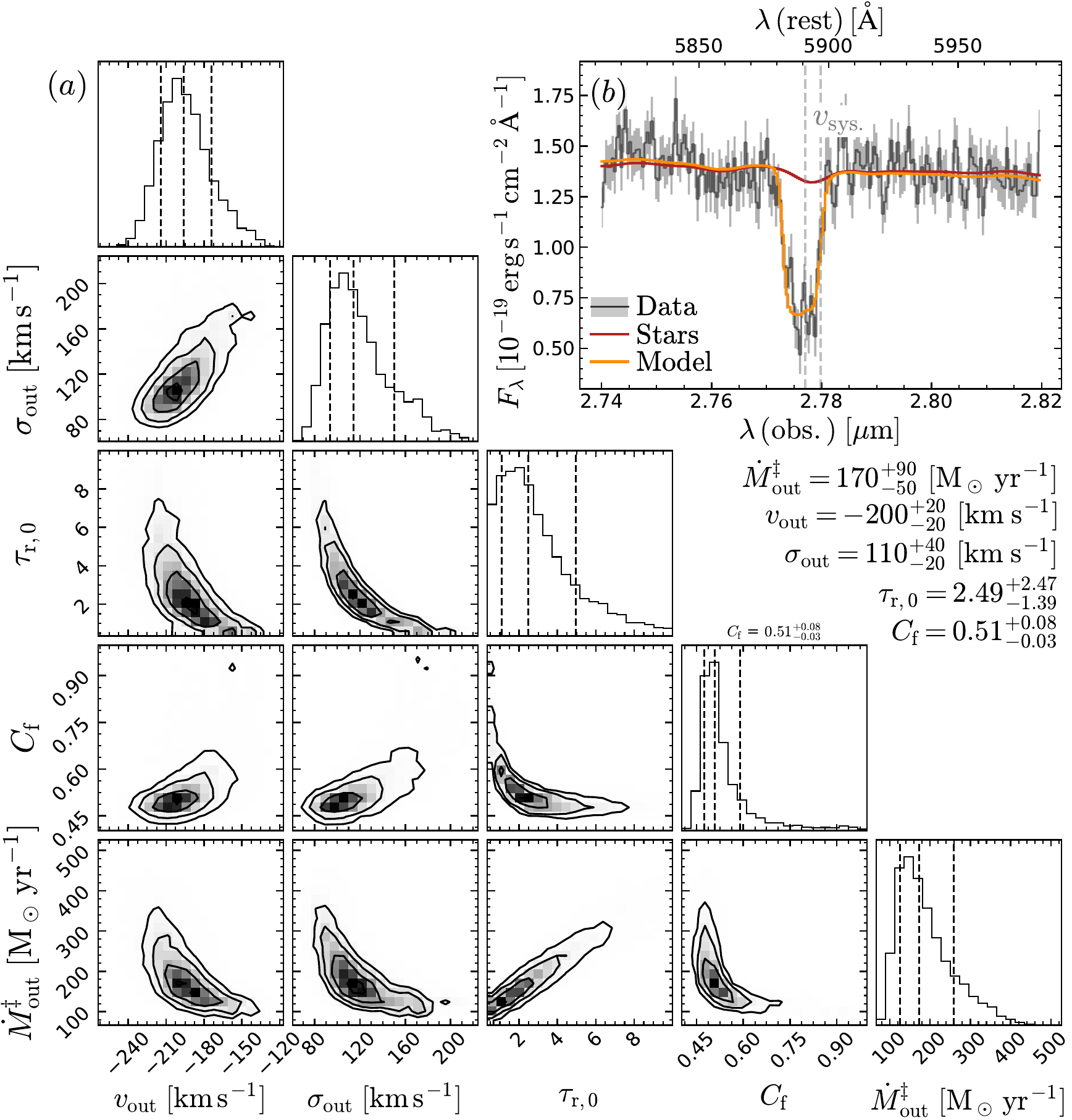}
\caption{{\bf Modeling the Na~I absorption in the foreground of the Jekyll galaxy as an outflow.} The corner plot (panel~a) shows the posterior distribution of the model parameters, with the contours in the 2-d
histograms representing the 0.5, 1, 1.5 and $2\sigma$ levels,
while the vertical bars in the 1-d histogram are the 16\textsuperscript{th},
50\textsuperscript{th} and 84\textsuperscript{th} percentiles of the
marginalized posterior distribution. The spectrum (panel~b) shows the fiducial model (orange), consisting of a stellar continuum (which includes stellar-atmospheric Na; red) and ISM absorption. If we assume the gas was an outflow driven from Jekyll, we estimate an outflow rate of $\dot{M}^\ddag_
\mathrm{out}=170_{-50}^{+90}\, \rm{M_\odot\,yr^{-1}}$.}\label{fig:nad}
\end{figure}

\begin{figure}
\centering
\includegraphics[trim=0.5cm 0.0cm 2.0cm 1.5cm,clip,width=1.0\textwidth]{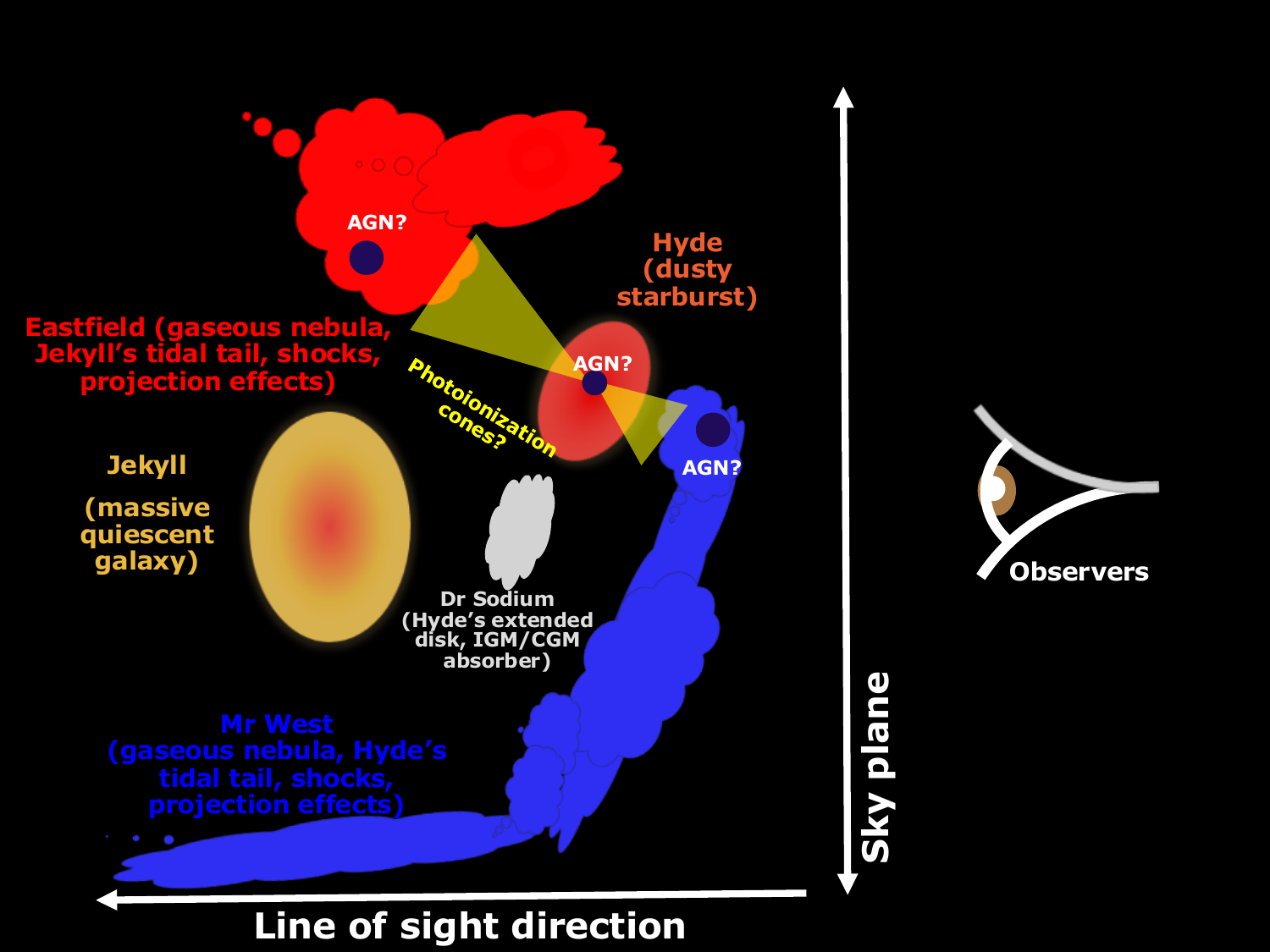}
\caption{\label{fig:cartoon}{\bf Cartoon of the {\it Tusitala} system.} The Hyde galaxy would be in the foreground with respect to Jekyll, given the existence of neutral gas likely associated with the dusty starburst. The position of the possible AGN present in the system are marked, and labeled with a question mark, as their existence cannot be confirmed with the currently available data.}
\end{figure}

\setcounter{figure}{0}


\clearpage
\section*{Methods}
\label{sec_methods}

\section{Spectroscopic and photometric data}
\label{methods:data}

This paper is mainly based on spectroscopy obtained within the JWST Guaranteed Time Observations (GTO) planned by the  Galaxy Assembly NIRSpec Instrument \cite{2022A&A...661A..82B} Integral Field Spectroscopy (GA-NIFS) Team (program ID 1217; PI: Nora L\"utzgendorf). It also used imaging data taken by the PRIMER program (ID 1837). We describe the main characteristics of these datasets in the following sections.


\subsection{JWST/NIRSpec integral field spectroscopic data}

JWST/NIRSpec observed the Jekyll \& Hyde system (coordinates J2000 10:00:14.76 $+$02:22:43.3) with its integral field unit on May 1st, 2023. We used  2 instrumental setups: prism low-resolution ($R=100$), wide spectral coverage (0.6-5.2~$\mu$m) spectroscopy, and grating observations high resolution ($R=2700$) coverage of the spectral ranges around the [OIII] and H$\alpha$ emission lines at $z\sim4$.

We downloaded the raw data files from the Barbara A.~Mikulski Archive for Space Telescopes (MAST) and processed them with the {\it JWST} Science Calibration pipeline version 1.8.2 under the Calibration Reference Data System (CRDS) context jwst\_1105.pmap. To increase the data quality, we made a number of modifications to the default reduction steps \cite[see][for details]{2023arXiv231003067P}. 
In brief, count-rate frames were corrected for $1/f$ noise through a polynomial fit. 
During calibration in stage 2, we removed regions affected by failed open MSA shutters, and by strong cosmic ray residuals in several exposures.
Remaining outliers were flagged in individual exposures using an algorithm similar to {\sc lacosmic} \cite{2001PASP..113.1420V}, following the procedure described by \cite{2023arXiv230806317D}: we calculated the derivative of the count-rate maps along the dispersion direction, normalized it by the local flux (or by three times the rms noise, whichever was highest), and rejected the 96\textsuperscript{th} (95\textsuperscript{th}) percentile of the resulting distribution for the prism (grating) data.
The final cubes were combined using the `drizzle' method, for which we used an official patch to correct a known bug (\url{https://github.com/spacetelescope/jwst/pull/7306}). The main analysis in this paper is based on the combined cubes covering a $3''\times3''$ area with a pixel scale of $0.06''$, we also checked the kinematics results and build the RGB map in Extended Data Figure~5 with a $R=2700$ $0.03''$ per spaxel cube to better identify the sub-kpc structures in ionized gas.

The $R=100$ cube was further prepared for stellar population synthesis analysis, beyond the pipeline reduction procedures. In particular, we matched the spatial resolution of each cube plane (i.e., each wavelength) to that corresponding to the reddest wavelength. We also tweaked the relative (as a function of wavelength) photometric calibration. For the first task, we used the PSF FWHM $vs.$ wavelength relationship published in \cite{2023arXiv230806317D}, which indicated that the spatial resolution degrades towards the reddest wavelengths, changing from $0.1''$ at $\sim1$~$\mu$m to $0.2''$ at $\sim5$~$\mu$m. We assumed a Gaussian PSF and applied a kernel to match the FWHM. Concerning the photometric calibration, we compared NIRCam broad-band fluxes measured with synthetic magnitudes obtained from the integrated spectra of 9 galaxies observed by the NIRSpec GTO GA-NIFS Team and reduced with the same pipeline and reference files versions, using the same photometric apertures. This exercise provided a fine-tuning calibration of the spectra, which was applied to the PSF-matched spectra for each spaxel. This correction varies from -5\% at $\sim1$~$\mu$m to +3\% at $\sim2$~$\mu$m, decreasing linearly to -5\% up to the reddest wavelengths.

The background for each wavelength of the cube as well as its noise used to estimate flux uncertainties were measured in two small circular apertures in the corners of the cube, which are $1''$ away from the galaxies. We multiplied the sky rms measured in the 0.06 arcsec/spaxel cube by 1.6 to account for the drizzling correlation.  The $R=100$ data reaches 5$\sigma$ detection limits of 27.2~mag per spectral element (approximately half of the spectral resolution) in the continuum around 1.3~$\mu$m (bluewards of the Balmer break for our galaxies, around the Mg$_\mathrm{UV}$ absorption feature for Jekyll) for a circular aperture of $0.1''$ radius (the FWHM of the PSF at these wavelengths), and 26.2~mag in the continuum around 3.0~$\mu$m (around the H$\alpha$ emission) for an aperture of $0.15''$ radius. For $R=2700$, the depths are 25.0~mag at 2.4~$\mu$m (around the [OIII] wavelength) for $0.1''$ apertures, and 24.5~mag for 3.1~$\mu$m (around the H$\alpha$ emission). These translate to line fluxes around  $3\times10^{-20}$~erg~s$^{-1}$~cm$^{-2}$ for the two mentioned wavelengths.


\subsection{JWST/NIRCam imaging data}

We made use of the NIRCam imaging of the Jekyll and Hyde system provided by the PRIMER program. We used images in 8 filters, namely F090W, F115W, F150W, F200W, F277W, F356W, F410M, and F444W with 5$\sigma$ depths measured in $0.3''$ diameter apertures of 27.6, 27.9, 28.1, 28.2, 28.3, 28.4, 27.7 and 28.0, respectively (in AB mag).

The data were reduced using the PRIMER NIRCAM imaging pipeline, called PENCIL (Magee et al., in preparation), which is built on top of STScI's JWST Calibration Pipeline (v1.12.5) but also includes additional processing steps not included in the standard calibration pipeline. For NIRCAM this includes the subtraction of $1/f$ noise striping patterns (both vertical and horizontal) that are not fully removed by the standard calibration pipeline and the subtraction of wisps artifacts from the short wavelength filters. During this processing we perform an additional step to identify and mask snowball artifacts that are not identified and masked during the calweb\_detector1 stage. Additionally, the background sky subtraction is performed by subtracting the median background over a grid while using a segmentation map to mask pixels attributed to sources and it automatically generates optimized source catalogs used for alignment. The image alignment is executed in two passes using the calibration pipeline’s TweakReg step and then using STScI python package TweakWCS: the first pass uses TweakReg to group overlapping images for each detector/filter and perform an internal alignment within the detector/filter group; the second performs alignment against an external catalog using TweakWCS. The external catalog was generated from an HST ACSWFC F814W image mosaic which was registered to the GAIA DR3 catalog. The final PRIMER image mosaics are generated using the calibration pipeline calweb\_image3 stage. In this final stage the pipeline also has the ability to split and process the data into smaller section when creating large mosaics and then stitch these smaller sections together to create a final full-size mosaic.

\subsection{ALMA dust continuum and nebular emission data}

\newcommand{ \CII} {[CII]}

We used ALMA band 8 observations of the \CII\ $\lambda$158~$\mu$m line from program 2015.A.00026.S  (P.I.  C.  Schreiber). 
 We obtained the calibrated measurement sets (MS) from the EU ALMA Regional Centre (ARC).  Data were calibrated with the standard pipeline procedure. 
To analyze the data, we used the ALMA data reduction software {\tt CASA} v6.2.1 \cite[][]{CASA}.

To create the data-cube of the \CII\ emission,  we subtracted the continuum from the two spectral window covering the line using the {\tt CASA} task {\tt uvcontsub},  assuming a constant continuum level. 
Then,  we generated the data-cube using the {\tt CASA} task {\tt tclean}, with Briggs weighting with robust parameter 0.5,  and cleaning threshold equal to  $3\times \text{RMS}$). The  \CII\ data-cube has a synthesized beam FWHM of $0.50\times 0.38$~arcsec$^2$.
We binned two spectral channels together to increase the S/N,  for a resulting channel width of $\sim25$~\kms. 
 The \CII\ data-cube has a sensitivity of $\sim 540$~$\mu$Jy~beam$^{-1}$ in a 25~\kms channel.

We extracted the \CII\ spectrum from an aperture of $0.25''$  centered on the position of the sub-mm continuum emission, and compared the velocity distribution with NIRSpec rest-frame optical measurements of the ionized gas (see Section~\ref{sec:el}). We also generated two  \CII\ channel intensity maps  by integrating the continuum subtracted data-cube in the spectral channels [-700,  -200]~\kms  and [-200,  100]~\kms with respect to $z=3.7135$.

We  created  a map of the continuum combining the two line-free spectral windows. 
To build the image,  we used the CASA task {\tt tclean}  applying  Briggs weighting with a robust parameter of 0.5.   This results in a synthesized beam with FWHM of $0.46\times 0.34$~arcsec$^2$. 
We used a cleaning threshold equal to  $3\times \text{RMS}$.
The continuum image has a sensitivity of $\sim42$~$\mu$Jy~beam$^{-1}$.  

We also made use of the high-resolution ($0.10\times 0.07$~arcsec$^2$ beam FWHM) ALMA Band 6 continuum data from project 2018.1.00216.S (P.I C.  Schreiber).  We used the continuum map provided in the ALMA archive,  reduced with the automatic pipeline, which has a sensitivity of $\sim15$~$\mu$Jy~beam$^{-1}$.
  
Extended Data Figure~1 shows the $\mathrm{A}(V)$ map built with our stellar population analysis in two dimensions  (see Section~\ref{methods:sps} and Figure~4) with the  ALMA contours overlaid. The ALMA dust continuum emission is located on top of the stellar mass density peak of Hyde, which present attenuation around 2.5~mag. The extension in the $\mathrm{A}(V)$ map to the NW overlaps with the most blueshifted [CII] emission and the extended continuum flux seen in the low spatial resolution map.

\section{Redshift determinations}
\label{methods:redshifts}

Due to the heterogeneous physical properties of the objects in the {\it Tusitala} group, we used different methods to calculate their redshifts. There is a known redshift bias between the prism and grating measurements \cite{2023arXiv230602467B}, likely due to residual calibration issues in the JWST data. Consequently, we favor the grating measurements over the prism results. 

\subsection{Jekyll's redshift}

For Jekyll, we added the spectra inside an aperture of semi-major axis 1~$R_\mathrm{e}$, and center, shape and position angle given by the 2\textsuperscript{nd} moment of the white cube image. Because Jekyll does not present any clearly detected emission lines, we used stellar kinematics. In Extended Data Figure~2, we show the analyzed aperture spectrum and the distribution of redshift $z$ and and aperture velocity dispersion $\sigma$ measured from 500 Monte-Carlo realizations of that spectrum, obtained by adding noise to the observed data. We measured $z$ and $\sigma$ using the same methods outlined in Section~\ref{methods:kinematics} below.

Our redshift estimation for Jekyll is $z_\mathrm{Jekyll}=3.7135\pm0.0002$,
which is inconsistent (3.5$\sigma$) with the previous value of $z=3.7174\pm0.0009$ \cite{2017Natur.544...71G,2018A&A...611A..22S}. We confirm our measurement is unchanged after testing different apertures (from $0.05''$ to $0.35''$) and stellar population libraries (see Section~\ref{methods:kinematics}). Extended Data Figure~2 panels a and b show the spectral region near the Balmer break; the red curve is our best-fit model, redshifted to $z=3.7174$; the stronger wiggles in the residuals rule out this solution for our data. If and how much the redshift difference is due to wavelength calibration offsets is still to be determined. For the rest of this article, we adopt the redshift of $3.7135$ for Jekyll and, given that it is the most massive member, for the whole {\it Tusitala} group.
The main results presented in this paper are not affected by this small redshift offset.

\subsection{Hyde's redshift}

For Hyde, we obtained the $R=2700$ spectrum from a circular aperture of radius 2.85~spaxels, centered on the object continuum. We do not detect H$\beta$ or [OIII], therefore we model only H$\alpha$ and [NII]. We take into account the instrument dispersion, integrate the Gaussian line profiles over each spectral pixel, and fix the [NII] doublet ratio to the value prescribed by atomic physics. The continuum is modeled as a straight line. This model is integrated using a Markov-Chain Monte-Carlo algorithm. The results are shown in Extended Data Figure~3, where we show the data and model, the residuals, and the posterior probability distribution of the most important parameters.
We marginally detect both H$\alpha$ and [NII] at 4-$\sigma$ significance. The resulting redshift $z=3.7076\pm0.0008$ is statistically consistent with the [CII] measurement of $z=3.7097\pm0.0004$ \cite{2018A&A...611A..22S}; the velocity dispersion $\sigma_\mathrm{ap}=220\pm40~\kms$ is consistent with the observed [CII] rotation velocity of $\sim200~\kms$ (Fig.~5 in \cite{2018A&A...611A..22S}).
We find a line ratio [NII]$\lambda$6583/H$\alpha=1.0^{+0.4}_{-0.3}$.


Due to the low detection level in the grating, we repeat our analysis with the prism data.
Here H$\alpha$ and [NII] emission is well detected but blended, so we model these lines alternatively by leaving their ratio free (orange lines and contours in Extended Data Figure~3), by using a Gaussian prior centered on the observed ratio and scatter of [NII]$\lambda$6583/H$\alpha$ measured from the grating (pink), and by fixing the line ratio (red). In all cases, we detect the line blend (8-$\sigma$ significance), but find a redshift $z=3.718\pm0.002$, higher than the grating value. However, the direction of the redshift difference is consistent with the known redshift bias between the prism and the medium-resolution gratings \cite{2023arXiv230602467B}. We calibrate this difference by measuring the offset between the [OIII] and H$\alpha$ emission in prism and grating data for the 2 regions in the field with the highest line fluxes. We derive an offset $\Delta z=0.0075$, and correct the prism redshift of Hyde accordingly, arriving to $z=3.711\pm0.004$. We remark that the detection of the line blend in a spectral region that is consistent with the grating measurement, and the fact that these lines have consistent fluxes, supports the fact that the H$\alpha$ detection in the grating is real, so we adopt $z_\mathrm{Hyde}=3.7076\pm0.0008$.

\section{Kinematic measurements}
\label{methods:kinematics}


\subsection{Measurements of Jekyll's kinematics based on the stellar continuum}

\subsubsection{Kinematics analysis of the integrated spectrum of Jekyll}

The stellar kinematics of Jekyll was obtained from the absorption spectrum detected in the $R=2700$ NIRSpec data. For estimating the aperture velocity dispersion, we used the same aperture as in the random-noise realizations we used for measuring Jekyll's redshift in Section~\ref{methods:redshifts}. We used the penalized PiXel-Fitting (pPXF) algorithm \cite{2004PASP..116..138C} and an initial redshift guess of $z=3.714$, with random initial conditions drawn from Gaussian distributions having mean and standard deviation of 0 and 200~km~s$^{-1}$ (for the velocity) and of 250 and 50~km~s$^{-1}$ (for the velocity dispersion). The continuum was modeled using a library of simple stellar populations (SSP) with the C3K model atmospheres \cite{2019ApJ...887..237C} and MIST isochrones \cite{choi16}, and spectral resolution $R=10,000$ over the rest-frame wavelength range covered by the NIRSpec high-resolution data. We mask the spectral regions near the edges of the detector (including the gap), and the prominent gas absorption feature (Ca~II$\lambda\lambda$3934,3968 at 3934.8 and~3969.6~\AA). Na~I$\lambda\lambda$5890,5896 was modeled as two Gaussians centered at 5891.6 and~5897.6~\AA, having the same velocity and velocity dispersion and line ratio between 1:1 and 2:1. We used a 15\textsuperscript{th}-order multiplicative Legendre polynomial to model out dust, possible flux-calibration issues, and systematic mismatch between the SSPs and the data.  The resulting redshift was discussed in Section~\ref{methods:redshifts}. 

The aperture velocity dispersion is
$\sigma_\mathrm{e}=261\pm15$~km~s$^{-1}$.
The most significant source of systematic error is the presence of an additive continuum, which is degenerate with $\sigma$. Significant residual background, emission-line infill and foreground continuum would all increase $\sigma$; if any of these were present, our measurement would be overestimated.
A strong continuum contamination seems ruled out by the fact that $\sigma$ varies by only 10\% between apertures of semi-major axis $0.05''$--$0.35''$. A significant contribution from Hyde is expected for the largest of these apertures, but Hyde contributes mostly at red wavelengths, whereas most of the $\sigma$ signal comes from rest-frame $B$-band features. Our model predicts no Balmer infill; if any was present, our $\sigma$ would need to be lower.
Our model requires weak [NII] emission to explain the profile of the H$\alpha$ absorption, similar but weaker than what is found for the quiescent massive galaxies described in \cite{2023arXiv230111413C}. These lines are present regardless of the aperture used, but their flux increases with the aperture radius, disfavoring a nuclear origin; their velocity is poorly constrained, but consistent with Jekyll within $\pm100~\kms$.
The origin of this emission could be linked to evolved stars, or to nebular emission around the Jekyll and Hyde system.

We tested that our results do not depend on the stellar library used; as alternatives to C3K/MIST, we used the XSL stellar library \cite{2022A&A...660A..34V} and E-MILES \cite{2016MNRAS.463.3409V}. For XSL, we masked the regions falling in the X-shooter dichroic; for E-MILES, we corrected the measured velocity dispersions empirically, by adding the E-MILES spectral resolution and subtracting the NIRSpec spectral resolution in quadrature from the measurement. Both libraries were converted from air to vacuum wavelengths.
Changing the aperture radius does not change $\sigma$, in agreement with the expectation that this measurement is almost independent of aperture radius \cite{2006MNRAS.366.1126C}

\subsubsection{Resolved kinematics analysis of Jekyll}

For the resolved kinematics, we follow the methods presented in \cite{2023arXiv230806317D}, with the only difference that we use a spaxel scale of 0.03~arcsec/spaxel. To obtain the necessary signal-to-noise ratio to study the stellar kinematics of Jekyll, we used Voronoi binning \cite{cappellari+2003} with a target signal-to-noise ratio of 15 per spectral pixel. We then fitted the spectrum of each bin using the same method as the integrated spectrum. Stellar rotation is clearly visible in the NIRSpec cube, extending to about $0.2''$ away from the center of Jekyll. For the stellar kinematics, we used only the bluest absorption lines (H$\gamma$ and bluer), which improved the signal-to-noise ratio of the velocity dispersion map. Fitting the entire spectrum results in a higher dispersion in the north half of Jekyll than in the center. The origin of this bias to higher dispersion is unclear, but the presence of Hyde suggests it could be due either to gravitational perturbation, or merely to the red continuum of Hyde contaminating the spectrum of Jekyll.
Regardless of the precise value of the velocity dispersion, Jekyll's kinematics and projected shape indicate a clear rotation pattern. This is consistent observations of a few other quiescent galaxies at lower redshifts $z=2\text{--}3$ \cite[e.g.,][]{2023arXiv230806317D}, as well as with the general expectation from other lines of evidence, including galaxy shapes \cite{2011ApJ...730...38V}, and statistical samples of galaxies at $z=0.5\text{--}1$ \cite{2021ApJS..256...44V,2023MNRAS.525.2789D}. A zoom-in version of the Jekyll stellar kinematics is shown in Extended Data Figure~4 (panels~a and~b); example spectral fits are shown in panels~e and~f.



\subsubsection{Detection of a neutral gas absorber in the foreground of Jekyll}

In addition to the stellar absorption, the spectrum of Jekyll flaunts unambiguous signatures of gas absorption along the line of sight, due to Ca~II$\lambda\lambda 3934,3968$ and Na~I$\lambda\lambda 5890,5896$. 
The non-stellar origin of these two features is evident from the large kinematic shift compared to the stars (Extended Data Figure~2 for Ca~II, and Figure~5 for Na~I).
We model Na~I as a set of two Gaussians, and repeat the spectral fitting procedure we used for the stellar kinematics. The resulting maps show that this absorption is everywhere blue-shifted with respect to Jekyll, and its velocity dispersion is significantly lower (only 200~\kms, or 40~percent lower in quadrature than 260~\kms). This is different from the properties of spatially resolved Na~I outflows, which show asymmetric spatial distribution and higher velocity dispersion \cite{2023arXiv230806317D}, and from the properties of unresolved outflows, which appear in conjunction with AGN-like nebular emission \cite{2024MNRAS.528.4976D}. Therefore, our fiducial interpretation is that this absorption originates from a cloud of gas (Dr Sodium) in the foreground of Jekyll. Assuming solar metallicity and a Milky-Way ionization fraction, we estimate a gas mass $\log\,M/\mathrm{M_\odot} = 8.4\pm0.2$, inversely proportional to the metallicity and to the neutral fraction. Given that Na~I absorption extends over the entire region where we detect Jekyll in the background, it is possible that our mass estimate for Dr Sodium is a lower limit.

Recent JWST NIRSpec/MSA observations reported that 46\% of massive galaxies at $z\sim2$ ($M_\star\sim10^{11}~\mathrm{M}_\odot$) display Na~I absorption in excess of the stellar expectations; this absorption is often blue-shifted (by less than 100~\kms for half of them), and attributed to neutral-gas outflows driven by AGN \cite{2024MNRAS.528.4976D}. We note that we do not find any evidence of nuclear activity in Jekyll; weak [NII] emission is spatially extended and has low EW.

Below, we repeat the analysis of Na~I absorption in Dr~Sodium with the same assumptions as \cite{2024MNRAS.528.4976D}, assuming that the absorbing gas is outflowing from Jekyll.
For Na~I, we model the aperture-integrated emission assuming an absorber with velocity $v_\mathrm{out}$ (relative to Jekyll's redshift) and dispersion $\sigma_\mathrm{out}$, Na~I$\lambda$5897 optical depth at the centre of the line $\tau_{\mathrm{r},0}$, and covering fraction $C_\mathrm{f}$. The optical depth of the bluest line is fixed to and $\tau_{\mathrm{b,0}} = 2\,\tau_{\mathrm{r,0}}$ \cite{2024MNRAS.528.4976D}. The stellar continuum absorption is fixed, by using the best-fit continuum model from pPXF. We also provide a 2\textsuperscript{nd}-order polynomial for adjusting the shape of the continuum. The likelihood function assumes uncorrelated Gaussian noise, and we provide flat priors on all parameters, with generous bounds. We calculate the posterior probability of this model using a Markov-Chain Monte-Carlo (MCMC) integrator \cite{2013PASP..125..306F}.
Figure~5 shows the posterior distribution (panel~a) and the maximum-likelihood model (panel~b). We find a clear velocity offset $v_\mathrm{out}=-200\pm15$~\kms, with a low dispersion of $\sigma_\mathrm{out}=114\pm5$~\kms. In agreement with expectations, there is a strong degeneracy between $C_\mathrm{f}$ and $\tau_{\mathrm{r},0}$.
The spatially resolved Na~I absorption reaches out to 1.8~kpc; if interpreted as an outflow with standard parameters \cite[solar metallicity and Milky Way ionisation fraction;][]{2024MNRAS.528.4976D},
this model would give a mass-outflow rate of $\dot{M}_\mathrm{out} = 170^{+90}_{-50} \mathrm{M^\ddag_\odot\,yr^{-1}}$ \cite{2024MNRAS.528.4976D}, where the $^\ddag$ symbol stresses that this is not our fiducial interpretation.


\subsection{Kinematics and excitation measurements based on emission lines}
\label{sec:el}

\subsubsection{Kinematics}

We perform the study of the ionized gas kinematics, the excitation mechanisms, and the properties of the interstellar medium through the spectral modeling of the strongest emission lines present in the NIRSpec IFU $R=2700$ data. To constrain the kinematics (line-of-sight velocity and velocity dispersion) of the ionized gas, we model the brightest line in our spectra [OIII]$\lambda$5007, keeping a fix ratio of 2.98 with the weaker [OIII]$\lambda$4959. This spectral modeling is performed at spaxel level in regions with $S/N > 3$ using a Gaussian profile for each line. This analysis provides the kinematics of the ionized gas in Eastfield and Mr~West; Jekyll and Hyde do not present [OIII] emission in the $R=2700$ data. The best-fit kinematic parameters were presented in Figure~2. Although these results assume a single kinematic component to represent the bulk motion of the gas, we note that our data provide both spatially and spectrally resolved components in the gas-rich system members. Some regions of Eastfield and Mr~West display complex line profiles indicative of high gas turbulence and/or the presence of ionized gas at different layers (velocities).

We further explore the distribution of the [OIII] emission through the velocity histograms presented in the left panel of Extended Data Figure~5, where we show, for comparison purposes, the distribution of the [CII]$\lambda$158$\mu$m emission. The zero-velocity corresponds to the redshift of Jekyll, $z=3.7135$. In the right panel of this figure, we show a three-color map highlighting the [OIII] and [CII]$\lambda$158$\mu$m emission at different velocities, as well as the emission from the stellar continuum (see caption for details). While the peak of [OIII] emission from Eastfield is consistent with the redshift of Jekyll, it presents a relatively broad (FWHM$ \sim $300-350~\kms) distribution of velocities that is slightly skewed towards negative values. The [OIII] emission from Mr~West is located at blue-shifted velocities and displays two peaks at -200~km/s and -750~km/s. This broad distribution of the [OIII] emission is similar to the wide [CII]$\lambda$158$\mu$m emission at the location of Hyde already identified by \cite{2018A&A...611A..22S}.

\subsubsection{Gas properties}

To constrain the excitation mechanisms in Eastfield and Mr~West, we employ the standard BPT optical diagnostic diagram \cite{1981PASP...93....5B}, which uses the line ratios [OIII]/H$\beta$ and [NII]/H$\alpha$. We explore the excitation mechanisms in small individual regions of $0.1''$ radius and in larger square regions encompassing each of the gas-rich group components, as shown in the right panel of Extended Data Figure~6. In these integrated spectra, we model all emission lines (H$\beta$, [OIII], H$\alpha$, [NII]) assuming the same kinematics in km~s$^{-1}$. For this analysis, we take into account the complex line profiles in some of the spectra by including a secondary kinematic component in the model, allowing us to recover the total flux across the full span of the line emission. The integrated line ratios obtained from this modeling are included in the BPT diagram presented in the left panel of Extended Data Figure~6. In the cases where [NII] or H$\beta$ present $S/N < 3$, we estimate the upper limits on their fluxes as $3\times$ the $RMS$ within the same FWHM as the other lines or, in the case of H$\beta$, as the most conservative between this estimate and the H$\alpha$ flux/3.0. Errors in the physical parameters derived in this paper are computed with MCMC techniques using the \textsc{emcee} software \cite{2013PASP..125..306F}, assuming uniform priors. 
The 1-$\sigma$ uncertainties on each parameter are calculated as half the difference between the 16th and 84th percentiles of the posterior probability density distribution. All our data points lie within the region consistent with AGN photoionization. In this diagram, we have also included grids and contours demarcating the line ratios that are consistent with ionization from shocks.

The mass of ionized gas in Eastfield and Mr~West are estimated based on their emission lines using standard properties (electron density and temperature $n_e=100$~cm$^{-3}$, $T_e=2\times10^4$~K), following \cite{2015ApJ...799...82C} and correcting for dust attenuation using the Balmer decrement. We find ionized gas masses $4.9_{-0.2}^{+0.4}\times10^8$~M$_\odot$ for Eastfield and  $4.3_{-0.3}^{+0.9}\times10^8$~M$_\odot$ for Mr West, with errors accounting for photometric uncertainties; systematic errors would be dominated by the unknown (and unmeasurable with our data) electron density. The integrated ionized gas mass of the gaseous component of the system is, therefore, $\sim 9\times 10^8$~M$_\odot$ each.



\section{Stellar population synthesis modeling}
\label{methods:sps}

The spectroscopic $R=100$ and $R=2700$ data were analyzed with three codes, {\sc synthesizer}, {\sc prospector}, and pPXF,  to obtain the physical properties of the stellar populations present in all the members of the galaxy group with a variety of approaches. We describe in the next three subsections the details and assumptions of the modeling for the three techniques and codes concerning Jekyll and Hyde, the two main galaxies in the system for which the continuum emission is very well detected. In Extended Data Figure~7, we show the SED fitting analysis.



\subsection{Analysis of the stellar populations in 2 dimensions with the {\sc synthesizer} code}

In order to analyze the stellar populations in Jekyll and Hyde separately, and given that they are blended to some extent, we first attempted a light decomposition of the NIRSpec spectra. To do that, we first fitted isophotes to the NIRCam data. We then translated those isophotes to the $R=100$ cube (wavelength by wavelength) an separated the emission of Jekyll from Hyde's, fixing the center and position angle to the results of the imaging analysis, and leaving the ellipticity free to account for the worse spatial resolution and distortions of the NIRSpec data. Spectra were eventually extracted for all spaxels for which more than 30\% of the full $R=100$ spectrum (sampled with 941 spectral pixels) have $S/N>3$. Each spectrum was fitted independently using the {\sc synthesizer} code, and then we merged the ones within the apertures shown in Figure~1 to characterize the global properties of the components of the system. The choice of data quality for each pixel (30\% of the spectrum with $S/N>3$) ensured that the region analyzed in 2D accounted for more than 90\% of the total flux of the system (enclosing Jekyll and Hyde) as measured in a large Kron aperture.

The {\sc synthesizer} code \cite{2003MNRAS.338..508P,2008ApJ...675..234P,2024arXiv240108782P} uses BC03 \cite{2003MNRAS.344.1000B} models including nebular emission as described in \cite{2003MNRAS.338..508P}, to simulate bursts described by a delayed-exponential Star Formation History, assuming Chabrier IMF \cite{2003PASP..115..763C} with stellar mass limits between 0.1 and 100~$\mathrm{M}_\odot$. The attenuation is modeled with the Calzetti law \cite{2000ApJ...533..682C}. We considered timescales $\tau$ between 30~Myr and 1~Gyr, metallicities between 0.2~Z$_\odot$ and solar, attenuations $\mathrm{A}(V)$ between 0 and 2 magnitudes for all galaxies except Hyde, for which we increased the range to 5~mag. The stellar population ages are allowed to vary from 1~Myr up to the age of the Universe at the redshift of the sources.

\subsection{Analysis of the stellar populations with integrated spectra using the {\sc prospector} code}

We extracted integrated prism spectra for the components of the {\it Tusitala} system and fitted them with the \texttt{Prospector} code in non-parametric mode. The \texttt{Prospector} (\cite{johnson21}, \cite{leja19}) is based on the Flexible Stellar Population Synthesis (FSPS;
\cite{conroy10}) library and uses the MIST stellar evolutionary tracks and isochrones \cite{choi16}, and the MILES stellar library to generate the stellar populations.  We used the following modeling assumptions for Jekyll: a \cite{2003PASP..115..763C} IMF, a stellar metallicity range between -1.5 and 0, and gas phase metallicity between -2.0 and 0.5. We included nebular lines and continuum emission generated using the \texttt{CLOUDY} \cite{cloudy}.  The ionization parameter for the nebular emission ranged between -4 and -1. The attenuation law used the two-component model which combines the \cite{cf00} two-component, birth-cloud vs. diffuse dust screens, approach with the \cite{kriek13} dust law where the UV dust bump and dust slope are correlated. Both components are modulated by a power-law factor that varies the slope of the attenuation relative to Calzetti. The ratio of
the nebular to diffuse attenuation ranged between 0 and 2, and the dust index of the power law ranges between -1 and 0.4. For the SFH we adopted a non-parametric model based on the flexible SFH prescription \cite{leja19} with 12 time bins and a continuity prior that controls the ratio of SFR in adjacent time bins. The first 3 bins sampled the recent SFH at lookback times of 5, 25, and 75 Myr, and the remaining 9 sample the rest at roughly equal intervals of $\sim$150~Myr. In addition to the stellar population properties, we allowed the redshift to vary within $\Delta z$/(1+z)=0.01 around z$=$3.71 to account for
small differences in the relative velocity of the different components of the {\it Tusitala} system.

We used small variations of the modeling assumptions for the SFH and dust attenuation to fit the SED of Hyde. Since Hyde is a star-forming galaxy and we are not aiming to determine a precise quenching time, we reduced the number of bins in the SFH to a typical prospector range of 6 bins to avoid additional uncertainties due to overfitting. The bins were spaced at 10, 100, 200, 500~Myr, and 1~Gyr to the age of the Universe at z=3.71. For the dust attenuation we adopted a simpler model based on a Calzetti law instead of the two-component model. Our initial runs based on the two-component dust attenuation yielded posterior distributions for $n$, the slope of the dust law, and the metallicity that were skewed toward the maximum grayness ($n\sim0.4$) and the lowest metallicity allowed by the priors. Since this could impact the reliability of the predictions in all the other parameters, we adopted a simpler model with fewer free parameters that leads to more regular posterior distributions. For Hyde, we also included in the fit the ALMA continuum measurement at $\sim750$~$\mu$m presented in \cite{2018A&A...611A..22S}. Prospector models the rest-frame IR emission using the \cite{2007ApJ...657..810D} dust emission templates characterized by three free parameters: U$_{\rm min}$ and $\gamma_{\rm e}$, which control the overall shape and peak of the IR emission and q$_{\rm PAH}$ which controls the strength of the PAH emission. The single ALMA point at $\sim$150~$\mu$m rest-frame does not probe the PAH region but it helps to estimate the bulk of the IR luminosity and the total dust attenuation through the energy balance constraints.  

\subsection{Analysis of the stellar populations in the core of Jekyll using the PPXF code}
\label{ppxf_jekyll}

The $R=2700$ data were used to estimate the ages and metallicities of the stellar populations in the core of the Jekyll galaxy (where the data were deep enough). We extracted spectra for elliptical annuli at different radii. Then each spectra was fitted using the MILES stellar population synthesis models \cite{2015MNRAS.449.1177V} and pPXF \cite{2004PASP..116..138C}. In short, pPXF returns the best-fitting linear combination of single stellar population models that best represent the input spectra. The weights of this linear combination can be then translated into mean ages and metallicities, as shown in Extended Data Figure~8. For the MILES models, we adopted the so-called base models, which inherit the chemical composition of the solar neighborhood, fed with BaSTI isochrones \cite{2006ApJ...642..797P}. We run pPXF imposing a regularized solution and a low-order multiplicative polynomial correction to the shape of the continuum. Neither of these assumptions has a significant effect on the recovered ages and metallicity values. Moreover, by construction, the quoted values for age and metallicity in the figure are mass-weighted. The ages obtained for the core of the Jekyll Galaxy using the $R=2700$ data are consistent with those obtained with the $R=100$ observations presented in Figure~4.



The NIRSpec data were also used to measure the magnesium abundance [Mg/Fe]. For that, we relied on the spectroscopic R = 100 data cube because of the NIRSpec gap in the Mg-sensitive region at ~5170~\AA\, (rest-frame). We measured the [Mg/Fe] over the central spaxels, using the same aperture as for the analysis of the high-resolution data. We followed an approach similar to that presented in \cite{2019A&A...626A.124M}, fitting every spectral resolution element within the standard Mgb line-strength definition (\cite{1994ApJS...94..687W}). We fixed the age and metallicity to the values derived from the high-resolution data and we simultaneously calculated the best-fitting [Mg/Fe] and effective spectral resolution using the Mgb line because assuming the nominal resolution led to poor fits to the data. For the fitting process we used the \cite{2015MNRAS.449.1177V} MILES models with variable [$\alpha$/Fe], linearly extrapolated above [Mg/Fe]=0.4 to match the observations. Our results are shown in Extended Data Figure~9. We obtain a relatively large [Mg/Fe] abundance, similar to that reported by \cite{2016Natur.540..248K} for a massive quiescent galaxy at lower redshift ($z\sim2$). The estimations for spectra extracted using different radii up to $0.16''$ consistently provide values [Mg/Fe]$ > $0.5. Similar results are also obtained when fixing the spectral resolution to 0.7 the nominal value, which has been shown to provide better results for the multi-object spectroscopic mode (NIRSpec JADES GTO Team, private communication).

\section{Morphological analysis}
\label{methods:morphology}

The NIRCam images, as well as the stellar mass map of Jekyll, were analyzed using the {\sc statmorph} software \cite{2019MNRAS.483.4140R} to study the morphological parameters of the galaxy. We used a Sérsic profile \cite{1968adga.book.....S} to fit the images in one and two dimensions. The photometric morphological analysis reveals that the galaxy is a spheroid with the following average parameters across all NIRCam bands redward of F200W: Sérsic index $n=3.0\pm0.7$ and effective radius $r_\mathrm{eff}=0.9\pm0.1$~kpc. This size is a factor of 1.7 larger than that presented in \cite{2015ApJ...808L..29S,2018A&A...611A..22S}, which was based on HST F160W data, where the galaxy is very compact, also confirmed by our F150W data, where we measure $r_\mathrm{eff}=0.5\pm0.1$~kpc. The JWST NIRCam data at $>2$~$\mu$m reveals a more extended galaxy. 

Concerning the mass profile, the central stellar mass surface density is $\sim2\times10^{10}$~M$_\odot$~kpc$^{-2}$, which is a factor of 2 higher than the values presented in \cite{2018A&A...611A..22S} using data with worse spatial resolution (thus averaging over a larger radius and resulting in smaller values).

Comparing with other massive quiescent galaxies at high-redshift, Jekyll is 35\% smaller (in terms of effective radius) than the system at $z=3.2$ presented in \cite{2024Natur.628..277G}; this galaxy would be even older than Jekyll (by $\sim0.5$~Gyr) assuming passive evolution between $z=3.7$ and $z=3.2$. Jekyll would be significantly more extended (a factor of $\times4$) than the $z=4.7$ galaxy discussed in \cite{2023arXiv230111413C}, which translates to a factor of five smaller central mass density; apart from an active nucleus, this galaxy presents a similar SFH timescale, a slightly younger age, which would translate to older ages for a passively evolving galaxy between the 2 redshifts. Finally, the properties of the $z=4.9$ massive galaxy in \cite{2024arXiv240405683D} are very similar to Jekyll's in terms of stellar and dynamical mass, morphology and SFH. All these examples provide a coherent picture of the early evolution of massive galaxies and its quenching, which for the first time we are able to probe with spatial resolution.


\section{Galaxy simulations}
\label{methods:tng}

We compared the mass accretion history of Jekyll with that of
massive galaxies in the suite of IllustrisTNG cosmological simulations
\cite[i.e., TNG50, TNG100, and TNG300;][]{2018MNRAS.475..676S, 2018MNRAS.477.1206N, 2018MNRAS.475..624N, 2018MNRAS.480.5113M, 2018MNRAS.475..648P}. In particular, we selected massive halos (dark matter masses $M_\mathrm{DM} > 10^{11.5}$~$M_{\odot}$) in the simulation and looked for massive galaxies
(stellar masses $M_{\star} > 10^{11}$~$M_{\odot}$) at redshift $z=3.7$ 
and followed the history of their progenitors.
This search led us to find six massive galaxies in TNG50,
26 galaxies in TNG100, and 299 galaxies in TNG300.
We noticed that most of these massive galaxies in the three simulations are not quiescent 
by $z=3.7$, building the bulk
of their stellar mass at later times than Jekyll (see Figure~3). Thus, in our analysis of the simulations, we concentrated in the most quiescent galaxies in IllustrisTNG, found mainly in TNG100 and TNG300. In particular, we focused our analysis in log($sSFR) < -0.30$~Gyr$^{-1}$ systems, finding six galaxies
in TNG100 and 23 galaxies in TNG300.


Among the selected galaxies in IllustrisTNG, ID~12071 (TNG300, $z=3.7$)
has the closest SFH compared to Jekyll, with two intense bursts
of star formation ($SFR\sim300$~$M_{\odot}$/$yr^{-1}$),
but delayed in time by 200-300~Myr.
In Extended data Figure~10, we show the mass accretion history
of both Jekyll and the 23 massive quiescent galaxies in TNG300, which highlights that Jekyll has an
accelerated life and death cycle compared to TNG300 systems (see also Figure~3).

We also looked for close massive ($M_{\star}>10^{10.5}~M_{\odot}$) companions of these massive quiescent galaxies, but found none within 20~kpc. Only one of the massive (but not quiescent) galaxies in TNG50 (ID~0, $z=3.7$) has a companion with stellar mass $\log(M_{\star}/M_{\odot})\sim10.5$ within 2.5~kpc.




\section{The {\it Tusitala} Group}

Figure~6 presented a cartoon representation of the system studied in this paper, where we we combine the spatial and spectral information from JWST to explore the line-of-sight direction. Here we describe the general properties of each component, providing supplementary information that supports and summarizes the analysis results in the article. Following \cite{2018A&A...611A..22S}, who named the galaxies Jekyll and Hyde from the novel {\it The strange case of Dr~Jekyll and Mr~Hyde} by Robert Louis Stevenson, we name the newly discovered components `Eastfield' (from the character Enfield), `Mr~West' (Mr~Guest), and `Dr~Sodium' (Dr~Lanyon). We call this system the `{\it Tusitala} Group', using the Samoan nickname of Stevenson, meaning `story teller'.

\subsection{Central galaxy: Jekyll}

The most massive component of the {\it Tusitala} Group is Jekyll.
The galaxy presents a spheroidal morphology and an effective radius typical of the cores of local elliptical galaxies \cite{2014MNRAS.444.2700D}. The stellar kinematics are complex, dominated by dispersion, but with a non-negligible rotation component. Using the calibration presented in \cite{2021ApJS..256...44V}, we infer a dynamical mass of $\mathrm{M}_{\mathrm{dyn}} =(0.9\pm0.2)\times10^{11}$~M$_\odot$ (defined as twice the mass within the effective radius). This is very similar to the stellar mass estimation (using a Chabrier initial mass function, \cite{2003PASP..115..763C}) obtained with pre-JWST data: $\mathrm{M}_{\star} =1.46-1.82\times10^{11}$~M$_\odot$ and $\mathrm{M}_{\star} =1.03-1.35\times10^{11}$~M$_\odot$ according to \cite{2017Natur.544...71G} and \cite{2018A&A...611A..22S}, respectively. It also matches well with the estimation based on our 2-dimensional SED fitting: $\mathrm{M}_{\star} =(1.16\pm0.15)\times10^{11}$~M$_\odot$ surviving stellar mass (i.e., not including remnants or winds) within a semi-major axis 3~kpc elliptical aperture with ellipticity 0.15. Accounting for a 0.07~dex difference between the stellar mass within 3~kpc and twice the mass within 0.85~kpc (the reference for the dynamical mass), we conclude that the ratio between the stellar mass and the dynamical masses is $\sim$1. This suggests that Jekyll has very little dark matter or gas within 0.85~kpc.


The accelerated development of Jekyll, resulting in a relatively evolved galaxy seen early in the history of the Universe, is demonstrated in Figure~3 and Extended Data Figure~10, where we provided its SFH. The most significant result is that the bulk of the stars formed rapidly in a very intense burst of star formation with a duration of 200-300~Myr, on the short end according to the integrated SED analysis, longer when accounting for some age and metallicity gradient obtained in our 2D approach. The process reached a SFR of $\sim400$~M$_\odot$~yr$^{-1}$ starting at $z\sim6$, 700~Myr before the epoch of observation. It quickly assembled $6\times10^{10}$~M$_\odot$ in $\sim200$~Myr (by $z\sim5$). Then, its star formation declined.
Jekyll is now effectively quenched, with a specific star formation rate $\mathrm{sSFR}=0.05\pm0.02$~Gyr$^{-1}$
and no signs of star formation or AGN, unlike other quiescent galaxies at high redshift (which present emission lines, e.g. \cite{2023arXiv230111413C,2024arXiv240405683D}). Based on this SFH, we infer a mass-weighted age for Jekyll of $650\pm50$~Myr.


The very quick SFH, that converted Jekyll in the massive quiescent galaxy we observe, also remarkably affected other two physical properties: the metallicity and $\alpha$-element abundances. From our low and high spectral resolution data independently, we infer that Jekyll's metallicity has a negative radial gradient, from nearly solar metallicity in the core, to around 0.5~Z$_\odot$ in the outskirts. The [Mg/Fe] abundance ratio is also high, $0.67\pm0.08$, consistent with the short formation timescale, and higher than typical nearby early-type galaxies of similar mass \cite{1994A&A...288...57M,2005ApJ...621..673T}. 


With these high metallicity values, the intense starburst where Jekyll formed must have also produced large amounts of dust (as in typical SMGs at $z\sim5-7$, \cite{2018Natur.553...51M,2023ApJ...946L..16P,2023ApJ...956...61A,2024arXiv240108782P}). However, the dust attenuation we observe in its current quiescent state is small, around $\mathrm{A}(V)=0.2-0.3$~mag and with a small positive radial gradient. This attenuation may not even be internal, but connected to Dr~Sodium foreground absorption (Figure~5 and see discussion below).


In Figure~3 and Extended Data Figure~10, we also demonstrated that at $z=3.7$, quiescent galaxies as massive as Jekyll are not reproduced by state-of-the art galaxy formation models, in our case illustrated with the results from Illustris \cite{2018MNRAS.475..676S, 2018MNRAS.477.1206N, 2018MNRAS.475..624N, 2018MNRAS.480.5113M, 2018MNRAS.475..648P, 2019MNRAS.490.3196P}. Leaving aside the early phases ($z\gtrsim8$) of the SFH, which are effected by large uncertainties in our analysis, the simulations typically predict a slower pace on the formation of massive galaxies compared to our results for Jekyll. Indeed, although some galaxies as massive as Jekyll are formed by $z\sim4$ in those simulations, and some of them present low sSFRs (around 0.1~Gyr$^{-1}$), they typically experienced their main star formation episode around 200~Myr, later than Jekyll. The average difference between the mass-weighted age of Jekyll and simulated galaxies is 150~Myr (with the total range spanning 140 to 220~Myr). We conclude that the Universe presents a faster evolution in the formation of massive galaxies compared to our more advanced simulations, implying the need for a better understanding about how baryons fall into the gravitational potential wells and transform into stars in the early Universe.

\subsection{Neighbor galaxy: Hyde}

Jekyll has a less massive but still heavy companion, Hyde, $\mathrm{M}_{\star} =(6\pm2)\times10^{10}$~M$_\odot$ measured within an aperture of 7.2~kpc diameter, separated by a projected distance of 2.9~kpc (0.4") and blueshifted by$\sim-375$~km~s$^{-1}$. If we interpret Hyde's [CII] emission as rotation (\cite{2018A&A...611A..22S}), considering that the velocity of the galaxy is -375~\kms\, (which is well located in the middle of the [CII] peaks in Extended Data Figure~5), its diameter is $\sim$2~kpc, and the [CII] emission tracing the disk has maximum velocity around 270~\kms, we obtain a dynamical mass $10^{10.5\pm0.4}$~M$_\odot$, considering an edge-on disk. This confirms the high-mass nature of the galaxy. 

Hyde is a dusty ($A_V\sim3$~mag), compact star-forming galaxy. Its morphology is typical of a disk, Sérsic index $n=1.0\pm0.1$, disk scale (effective radius in a Sérsic profile) $r_\mathrm{eff}=2.4\pm0.8$~kpc. The morphology is dominated by an unresolved clump but presents a faint extension to the Northwest, as shown in the dust attenuation map in Figure~4. This galaxy is experiencing an intense burst of star formation in the last 100--150~Myr, reaching SFRs as high as 300~M$_\odot$~yr$^{-1}$ (based on the stellar population analysis and also directly on the detection of H$\alpha$ emission in the NIRSpec $R=100$ data, corrected for dust extinction), similar to what was reached by Jekyll at the peak of its SFH. Both the 2-dimensional and integrated spectral energy distribution analyses indicate that Hyde is already quenching (as suggested by \cite{2021A&A...650C...2S}). Based on the SFHs shown in Figure~3, we infer a mass-weighted age for Hyde between 100 and 300~Myr.

\subsection{Gas-rich components of the group: Eastfield, Mr~West, and Dr~Sodium}

The NIRSpec IFU data uniquely reveal two other components of this system that are aligned with Hyde: Eastfield and Mr~West. The direction defined by Eastfield, Hyde, and Mr~West, is roughly perpendicular to the rotation inferred from the [CII] map.
 
These sources are dominated by emission lines, extending over 11~kpc (1.6") projected distance on either side of Hyde (i.e., 22 kpc, edge to edge). The filamentary western component presents clumps blueshifted by up to $-750$~km~s$^{-1}$ relative to Jekyll, but also significant emission is blueshifted by just $-200$~km~s$^{-1}$. In contrast, Eastfield is more extended and diffuse, with a more symmetrical velocity distribution around the reference set in Jekyll, with around half of the emission blueshifted and half redshifted by up to $\sim200$~km~s$^{-1}$ (Extended Data Figure~5). Our data do not constrain the stellar mass of these two systems, because the nebular continuum emission from both Eastfield and Mr~West is very strong and could account for most of their observed luminosity. In any case, our upper limit
imply both components are significantly less massive than
Jekyll and Hyde (see Methods).

The velocity distribution of the Eastfield and Mr~West structures seen by JWST in H$\alpha$ and [OIII] correlates well (see Extended Data Figure~5) with the velocity from [CII] (\cite{2018A&A...611A..22S}), all exhibiting several emission peaks concentrated in the velocity range between -1,000 and +200~km~s$^{-1}$ with respect to Jekyll. However, the spatial location is completely different, the [CII] emission being concentrated in Hyde, and the optical emission lines mainly arising from Eastfield and Mr~West.

The velocity map and alignment with Hyde of the two gas-rich regions point to a filamentary structure of orbiting gas that was part or was probably feeding the starburst in Hyde. Indeed, their velocity and structure are consistent with tidal tails along the trajectory followed by Hyde and Jekyll in their interaction, or with intergalactic clouds being perturbed and heated by the star formation and, most probably, nuclear activity of the massive galaxies. In particular, Mr~West presents a wide range of velocities which perfectly match what is expected for a tidal tail \cite{1972ApJ...178..623T,2013LNP...861..327D} following the trajectory of Hyde in the encounter with Jekyll (which eventually can end up in a merger, as the $\Lambda$CDM paradigm predicts). The nature of Eastfield is less clear, but its alignment with Hyde and Mr~West, its clumpy structure within an extended circular shape, the large dispersion seen in some of the clumps, and the emission-line ratios, all point to some strong energy source being responsible of its heating and ionization state. This strong energy source could be an active nucleus in Hyde (or within the cloud itself, as proposed by \cite{2023arXiv231003067P}) photoionizing (and shocking, as seen in Hanny’s Voorwerp object, \cite{2009MNRAS.399..129L}) the circum-galactic material in Eastfield.



Finally, the NIRSpec IFU data also reveal another ingredient of our current galaxy formation paradigm: the interaction of galaxies with their environment through the enrichment of the circum- and inter-galactic medium via strong winds produced by star formation or nuclear activity, which eventually provoke star formation quenching. Two proofs are offered by the JWST data of this interaction of galaxies with their neighborhood. First, the high signal-to-noise spectrum of Jekyll presents very strong NaD absorption offset (blueshifted) by $-$200~km~s$^{-1}$ from Jekyll, which indicates a non-stellar origin and most probably not linked to the interstellar medium of the massive quiescent galaxy, but to the circumgalactic medium. We call this component of the system Dr~Sodium. This absorbing material has a larger (blueshifted) velocity with respect to Jekyll compared to that of Hyde. Given that Jekyll is a dead galaxy, and the velocity of Dr~Sodium is large, $-$200~km~s$^{-1}$, the most probable origin of the enriched gas in Dr~Sodium is Hyde. The possible explanations are: (1) An extension of the disk seen in [CII], since Dr~Sodium extends in the same Southeast direction of the ionized nebular line and with similar velocity. (2) Tidally removed material from the massive galaxies, as also inferred for Mr~West. (3) In a similar scenario, the NaD absorption could be linked to Mr~West, Dr~Sodium  would then be an extension of that gas-rich system, whose origin point to a tidal tail from Hyde. This is the first time that strong signal from neutral gas is detected around a dusty star-forming galaxy (with no indications for the presence of an AGN) at high redshift, a phenomenon that has been observed in the nearby Universe \cite{2016A&A...590A.125C}.

The second proof of the interaction of these massive galaxies with their environment, affecting a region at the same distance of the NaD intergalactic medium absorption ($\sim3$~kpc), is the large dispersion, $213$~km~s$^{-1}$, observed in the emission lines arising from the regions in Eastfield and Mr~West which are closest to Hyde (see Figure~2). These dispersions can be interpreted as turbulence induced by the AGN/starburst outflows, maybe interacting with in-flowing material which would be feeding the star formation in Hyde. An alternative explanation can be the integration along the line of sight over a wide range of velocities arising from a filamentary structure orbiting the galaxies.

And the final possible phenomenon that might explain the kinematics of the gas is the presence of shocks whose origin can be star formation and/or, most probably, nuclear activity in Hyde, which is experiencing a strong starburst (typically also activating the SMBH, \cite{1988ApJ...325...74S}), Jekyll (whose quiescent nature might rather be linked to a radio-mode active nucleus, e.g., \cite{2006MNRAS.365...11C}), or even an active nucleus within Eastfield and Mr~West themselves (as proposed by \cite{2023arXiv231003067P}).


A possible third proof of the presence of strong outflows from Hyde is the dust distribution, revealed by the A(V) map in Figure~4 and especially the dust continuum and [CII] emission obtained with ALMA \cite{2018A&A...611A..22S} presented in Extended Data Figure~1. In particular, the [CII] shows a large velocity component to the NW, in opposite direction of the NaD absorption.

\subsection{Where are the AGN in the {\it Tusitala} group?}


The emission-line ratios measured for the entire Eastfield and Mr~West group components, as well as for individual regions within these clouds, are inconsistent with photoionization due to star formation (Extended Data Figure~6),
leaving AGN photoionization and shocks as the only possible
explanations. In principle, shocks can be ruled out as the dominating excitation mechanism by the high observed luminosity \cite{2023arXiv231003067P}, but we remark that the
conditions explored by \textsc{mappings\,v} may not be
representative of the large scales of this system.
Assuming typical AGN line ratios apply to $z\sim4$ and diffuse sources such as Eastfield and Mr~West, the main ionizing source should be nuclear activity linked to the two massive galaxies Jekyll and Hyde, and/or AGN in the gas-rich low-mass members of the system, Eastfield and Mr~West. This, however, does not rule out some contribution from shocks, which may also increase the density in some specific regions of Eastfield and Mr West, leading to small star-formation knots, as seen in the local Universe \cite{2009MNRAS.399..129L,2012AJ....144...66K,2023A&A...678A.127V}. Brighter clumps are indeed seen in the regions of Eastfield and Mr~West which are closest to Hyde; we calculate the stellar masses of those clumps to be around $10^8$~M$_\odot$ and smaller, with large uncertainties due to the very faint continuum which in part can have a nebular origin. The integrated stellar masses of the Eastfield and Mr~West systems are around $10^9$~M$_\odot$ each, but these values should be considered as upper limits. 


With the current data, we find no proof of the presence of one or several AGN in the system. However, the limited observations cannot completely rule out the presence of AGN in some locations of the {\it Tusitala} group.

One of the  possibilities is that of an active nucleus is Hyde, which is forming stars very intensely in a dusty burst. In this scenario, both Eastfield and Mr~West would fall in the AGN ionization cones, thus explaining the presence of AGN-like line ratios on either side of Hyde; our discovery  of non-emitting neutral gas in the foreground of Jekyll could be readily explained by Dr~Sodium being outside the AGN cone (Figure~6). Moreover, the [CII] velocity field, interpreted as rotation in \cite{2018A&A...611A..22S}, and roughly perpendicular to the line joining Eastfield, Hyde and Mr~West, would be also consistent with this scenario.

The presence of an accreting SMBH in Hyde is plausible, even though we cannot detect it with the currently available data. The region where the dust continuum peaks is very compact ($<2$~kpc diameter, see Extended Data Figure~1) and its stellar mass is close to $\sim10^{11}$~M$_\odot$. A SMBH of around $10^7$~M$_\odot$ could be present in that region \cite{1998AJ....115.2285M}. Based on the H$\alpha$ emission in the nebulae (assuming it is entirely due to AGN photoionization), and their angle subtended to Hyde, we can infer the luminosity that the AGN in Hyde must have (or must have had, if faded) to explain the observed H$\alpha$ luminosity.
H$\alpha$ is a simple recombination line, hence its luminosity is primarily a `photon counter' and  largely independent of ionization parameter and metallicity. By using the broad range of AGN ionizing spectral shapes explored in \cite{Nakajima2022} we obtain that, for a medium entirely covering an AGN, the ratio between H$\alpha$ luminosity and AGN bolometric luminosity is in the range between $10^{-2}$ and $10^{-3}$. Hence, for a cloud that covers less than 4$\pi$ this ratio is reduced by an amount corresponding to the covering factor. In the case of Eastfield and Mr~West, we estimate projected covering angles $\theta$ relative to Hyde of $\sim 48^\circ$ and $\sim 35^\circ$, respectively. We translate these in covering factors by simply assuming $C_\mathrm{f} \approx  \frac{\left( \theta ~(\pi /180)\right) ^2}{4\pi}$. From this, Eastfield would give that the putative AGN in Hyde should have (or had, in the past) a bolometric luminosity in the range $\log(L_\mathrm{AGN}/(\mathrm{erg\,s^{-1}}))\sim 45.2\text{--}46.2$, while Mr~West would give $\log(L_\mathrm{AGN}/(\mathrm{erg\,s^{-1}})) \sim 44.5\text{--}45.5$. These calculations assume that the projected distance (hence projected angles) are the real distances and that each of the two clouds has internally a covering factor of unity (i.e. not fragmented in smaller clouds), assumptions which lead to substantial uncertainty. This can be compared with the upper limit on the 24~$\mu$m emission measured by Spitzer-MIPS, corresponding to 5.1~$\mu$m rest-frame.
If we adopted a standard, type 1 SED \cite[as measured by][]{Maiolino2007}, this would imply a $\log{(\nu L_\nu)}$ luminosity at 5.1~$\mu$m  in the range 44.2--45.9~erg~s$^{-1}$. This would correspond to a MIPS flux between 22-1100~$\mu$Jy, mostly inconsistent with the observed 5$\sigma$ upper limit of $\sim100$~$\mu$Jy (\cite{2019ApJS..243...22B}). However, in type 2 AGN, especially Compton-thick ones, the 5~$\mu$m to bolometric ratio can be much lower (also in hot or warm dust deficient quasars, \cite{2017ApJ...835..257L}). For instance, \cite{Stalevski2017} model the IR SED of the Circinus nucleus, a local, prototypical type 2 AGN, by obtaining $L_{bol}/[\nu L_\nu]_{5\mu m}\approx 40$. If we adopt the same model for the presumably equally obscured AGN in Hyde, we would obtain that the flux at 24~$\mu$m should be in the range 1--60~$\mu$Jy, so compatible with the MIPS upper limit.
Note that these calculations apply for an AGN that is active at the time of observation. If the AGN has faded and the nebulae are only the echo of the past activity (as in Hanny's Voorwerp object, \cite{2009MNRAS.399..129L}), then the MIPS upper limit is obviously not constraining.

Finally, Jekyll could also host an AGN. Its quiescent nature would favor a radio-mode AGN similar to those found in nearby elliptical galaxies \cite{1974MNRAS.167P..31F}. However, its old mass-weighted age and the presence of a more active companion, Hyde, or companions, would favor a small role of Jekyll in the excitation of the gas in its surroundings in the most recent times. In any case, at the location of Jekyll, we can rule out bright radio emission, with the 3~GHz flux being $<7~\mu$Jy \cite{2017A&A...602A...2S}.

\clearpage

\renewcommand{\figurename}{Extended Data Figure}
\begin{figure}
\centering
\includegraphics[trim=0.0cm 0.0cm 0.0cm 0.0cm,width=0.3\textwidth]{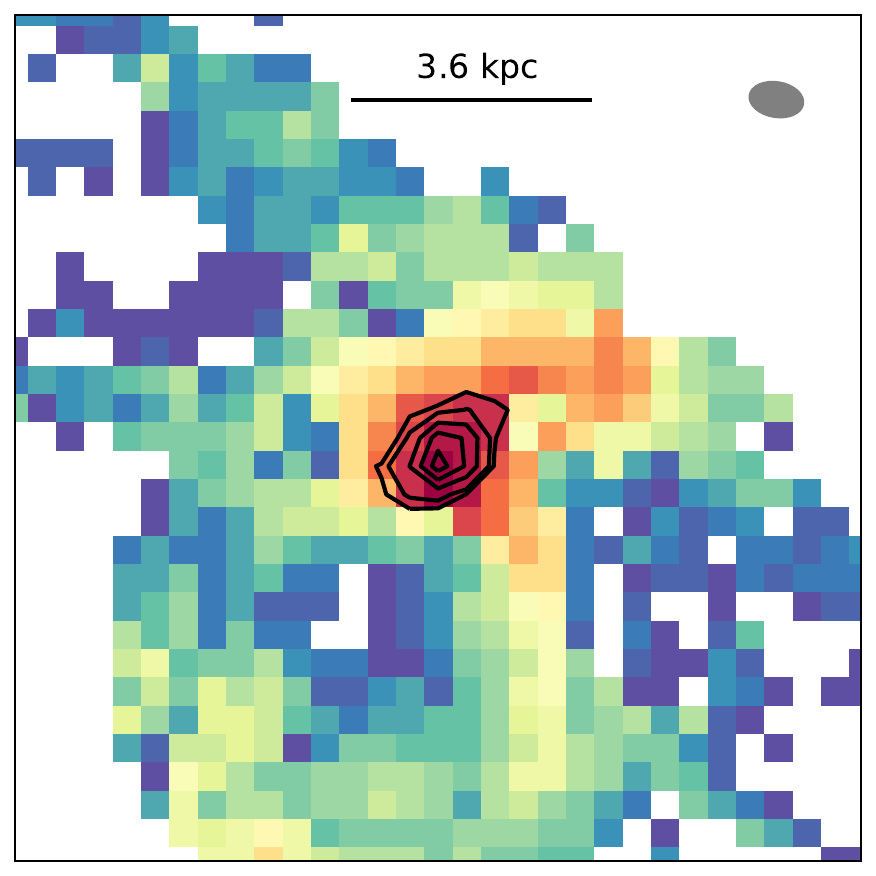}
\includegraphics[width=0.3\textwidth]{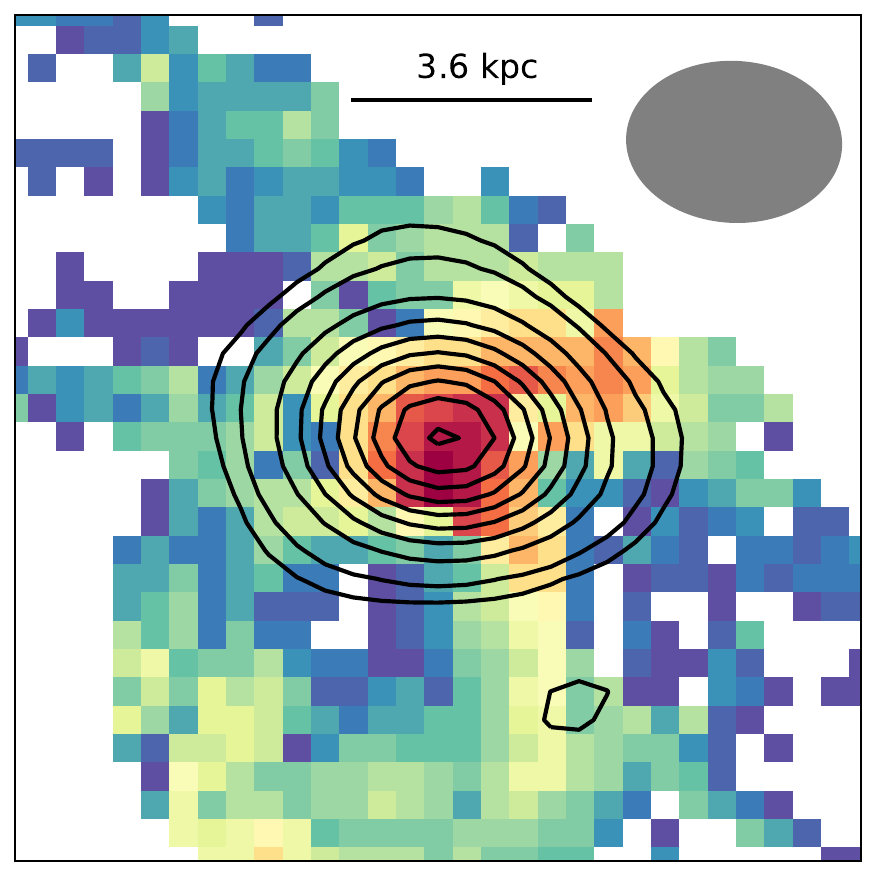}
\includegraphics[width=0.37\textwidth]{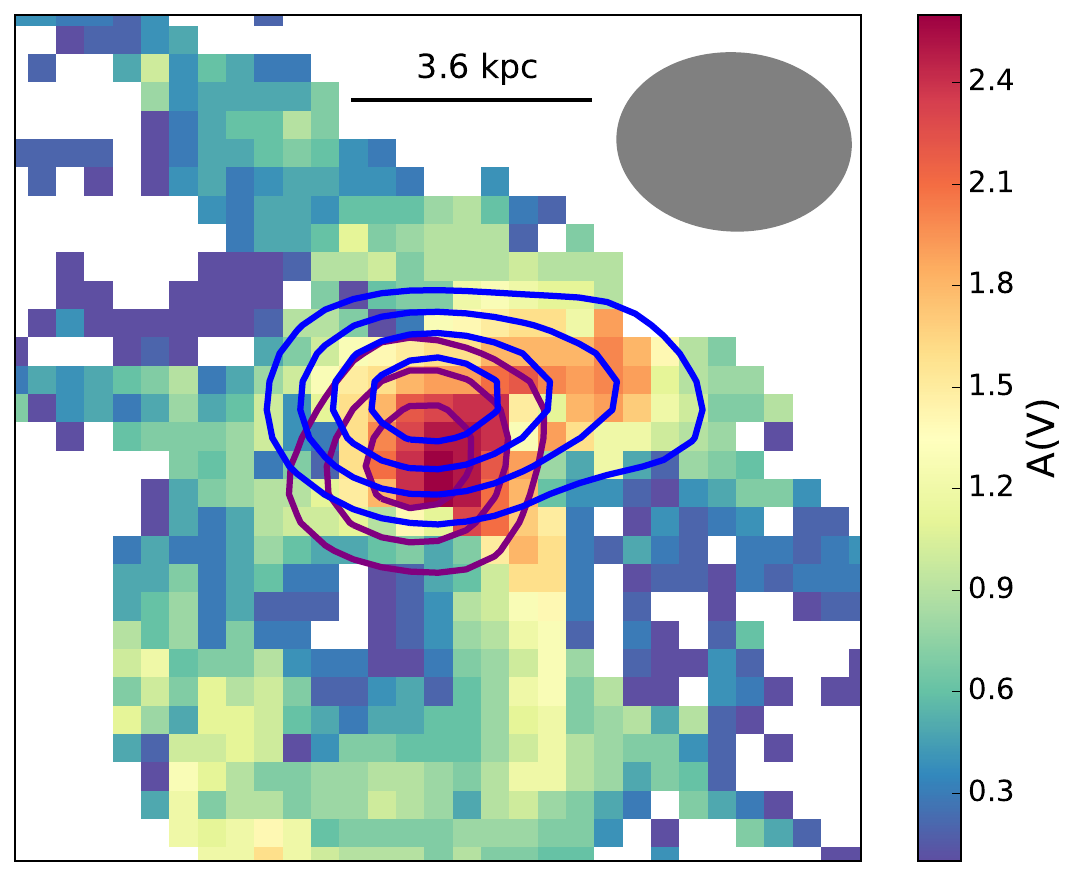}
\caption{{\bf Attenuation maps from the NIRSpec IFU stellar population analysis.} Jointly with the $\mathrm{A}(V)$ maps, we  overlay ALMA contours from:  continuum high-resolution (beam FWHM $=$$0.10\times0.07$~arcsec$^2$) observation at 240~$\mu$m rest-frame (left), continuum low-resolution (beam FWHM $0.46\times0.34$~arcsec$^2$) at 158~$\mu$m rest-frame (middle), and \CII\ $\lambda$158~$\mu$m emission in two velocity channels ([-700, -200] \kms in blue and [-200, 100] \kms in red) (right). Continuum contours are at [3,5,10, 15, 20, 25,30, 35]$\times\sigma$. 
[CII] contours start at 3$\sigma$ and increase by 2$\sigma$. 
The FWHM and position angle of the ALMA beams are indicated by the  gray ellipses.}
\label{fig:AV_ALMA}
\end{figure}

\begin{figure}
\centering
\includegraphics[width=1.00\textwidth]{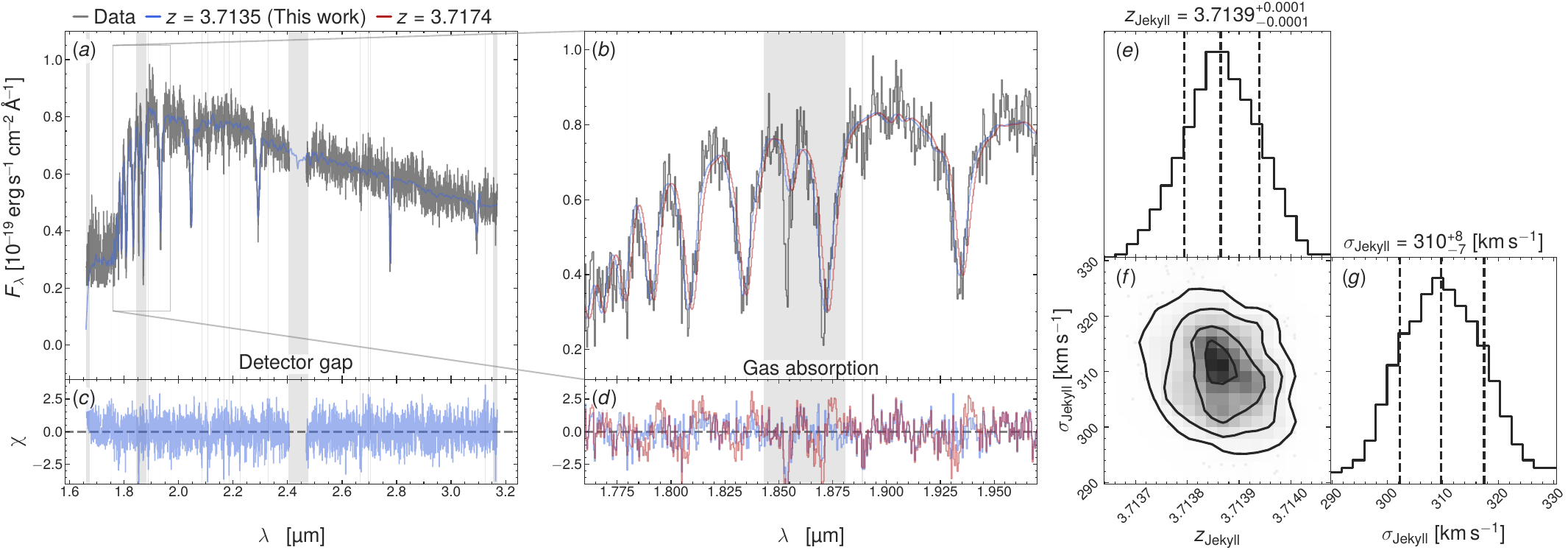}
\caption{{\bf Redshift and dispersion of the Jekyll galaxy.} Left: a. NIRSpec high-resolution spectrum of Jekyll, obtained by integrating inside an elliptical aperture of semi-major axis $0.25''$, 1.8~kpc (gray). Our best-fit pPXF model is shown in blue, giving a redshift $z=3.7135\pm0.0002$. b. Detail of the spectral region around the Balmer break, highlighting in red a solution at $z=3.7174$ \cite{2017Natur.544...71G,2018A&A...611A..22S}, which our JWST data rule out. Panels c and~d show the residuals, normalized by the noise. Vertical gray bands have been excluded from the fit. Right: e. Distribution of redshift $z$ and aperture dispersion $\sigma_\mathrm{ap}$ for five hundred Monte-Carlo realisations of the Jekyll spectrum. f. Histrogram of solutions for redshift. g. Histogram of solutions for dispersion. The contours in the 2-d
histograms represent the 0.5, 1, 1.5 and $2\sigma$ levels,
while the vertical bars in the 1-d histogram are the 16\textsuperscript{th},
50\textsuperscript{th} and 84\textsuperscript{th} percentiles of the
marginalized posterior distribution. The reported
values are
the median and the inter-percentile range of the marginalized posterior probability distributions.
}\label{fig:redshift}
\end{figure}

\begin{figure}
 \includegraphics[width=1.\textwidth]{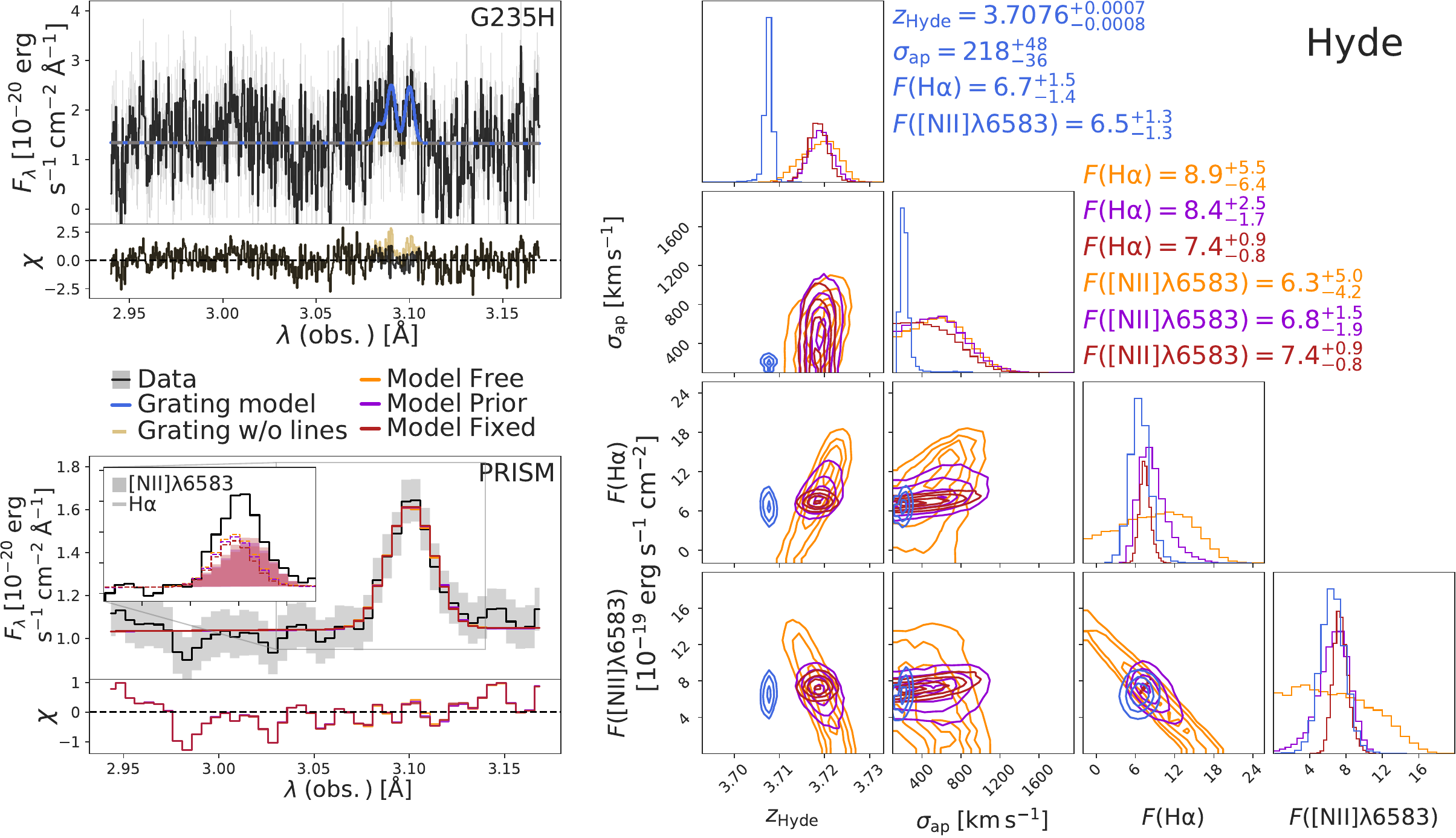}
 \caption{\label{fig:redshifts.hyde}{\bf Redshift determination for the Hyde galaxy.} The high-resolution grating (top left; blue lines) has only a marginal detection of H$\alpha$ and [NII]$\lambda\lambda$6548,6583 (4-$\sigma$ significance each), at a redshift of $z_\mathrm{Hyde}=3.7076$ that is consistent with the ALMA [CII] measurement. In the prism data (bottom left), H$\alpha$ and [NII] are very prominent but blended. We find a clear detection of this blend, with or without using prior information from the grating (the orange, pink, and red lines are three alternative models). The corner diagram on the right shows the posterior distribution of the grating and prism models. The contours in the 2-d
histograms represent the 0.5, 1, 1.5 and $2\sigma$ levels,
while the vertical bars in the 1-d histogram are the 16\textsuperscript{th},
50\textsuperscript{th} and 84\textsuperscript{th} percentiles of the
marginalized posterior distribution. The reported
values are
the median and the inter-percentile range of the marginalized posterior probability distributions.
 The flux measurement is consistent between the two dispersers; the redshift is not consistent, but admitting that this could be to known wavelength calibration issues between the NIRSpec dispersers, we calibrate the difference using bright emitting regions in the system.}
\end{figure}

\begin{figure}
    \centering
    \includegraphics[width=\textwidth]{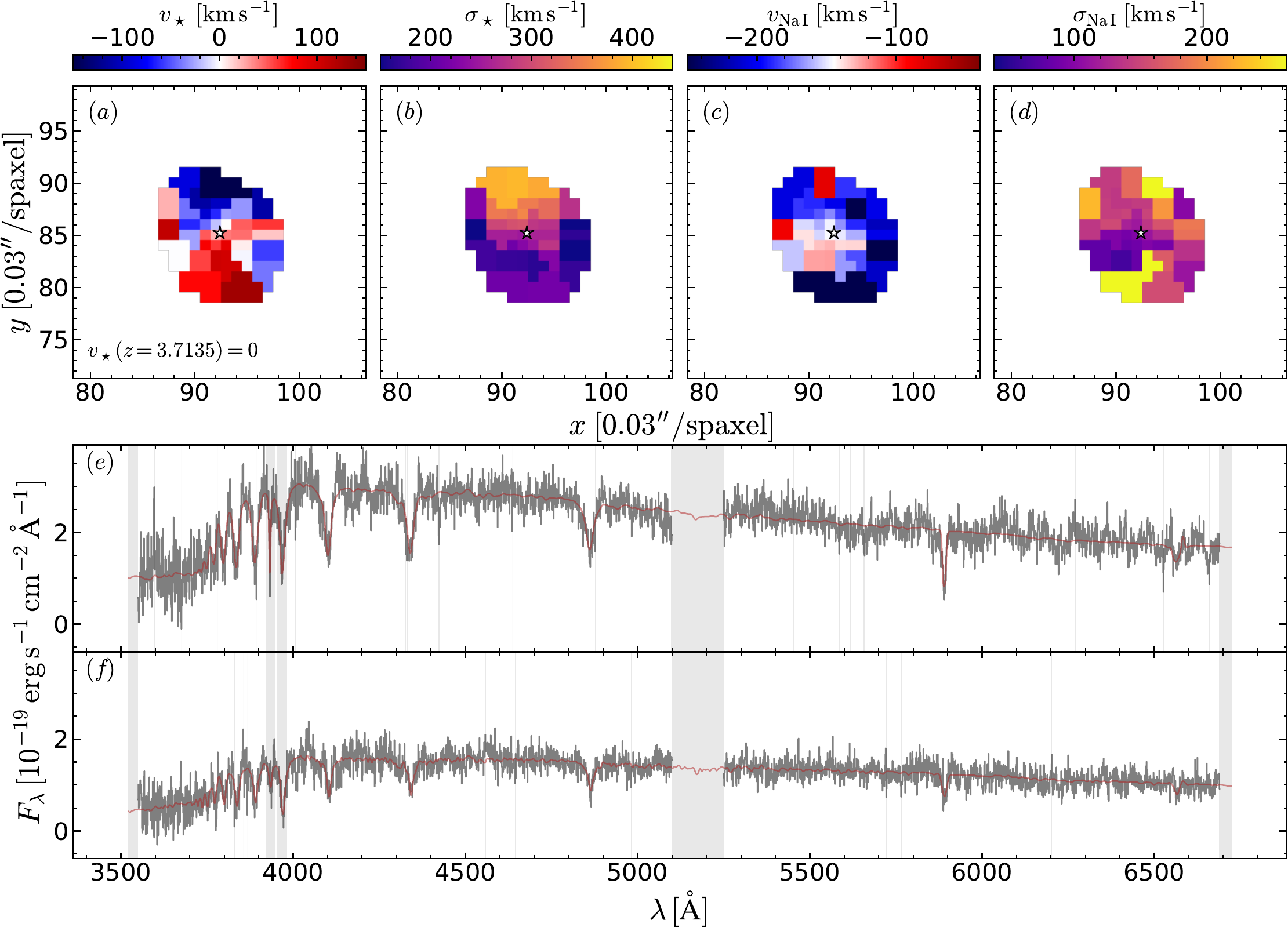}
    \caption{{\bf Maps of absorption kinematics for Jekyll.} We show stellar velocity and velocity dispersion (panels~a and~b), and Na~I velocity and velocity dispersion (panels~c and~d). Panels~e and~f show two example spectra from two spaxels on opposite sides along the major axis of the galaxy.
    }
    \label{fig:jekyllmaps}
\end{figure}

\begin{figure}
\centering
\includegraphics[trim=0.0cm 0.4cm 0.0cm 0.0cm,clip,width=0.47\textwidth]{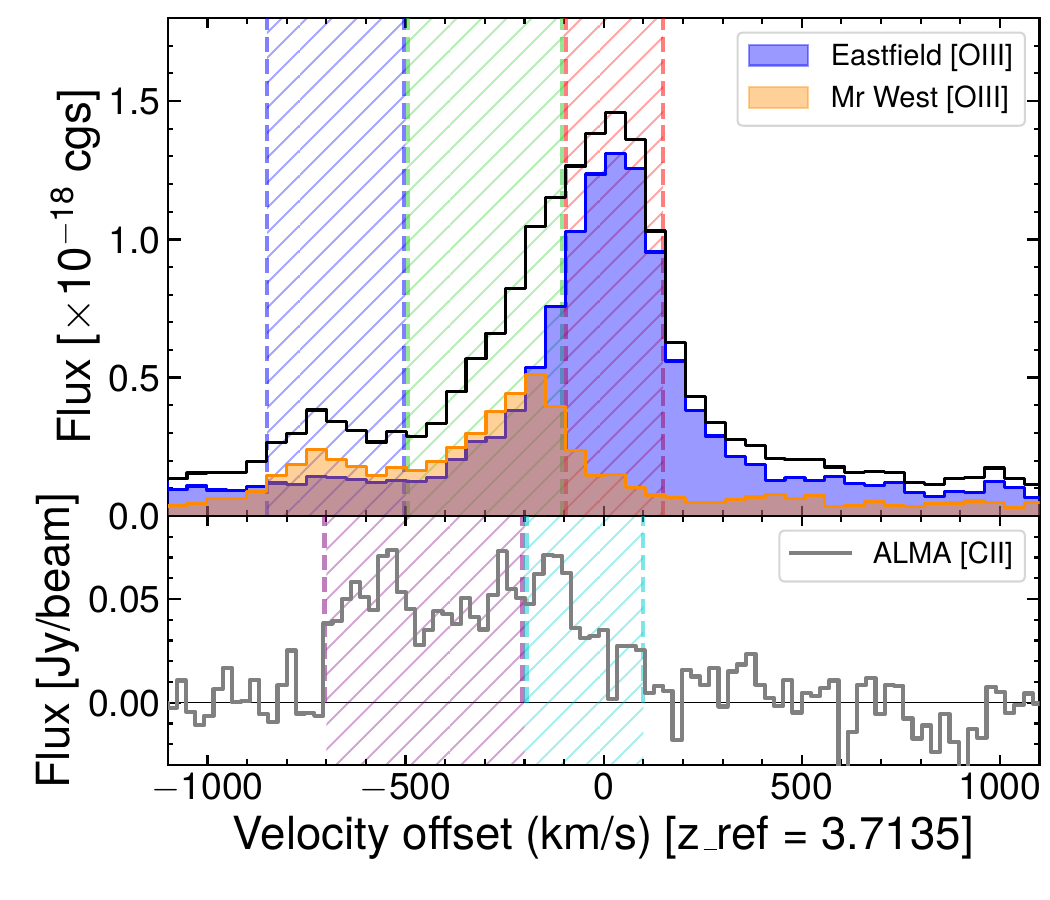}
\hspace{0.2cm}
\includegraphics[trim=0.3cm 0.3cm 0.3cm 1.0cm,clip,width=0.48\textwidth]{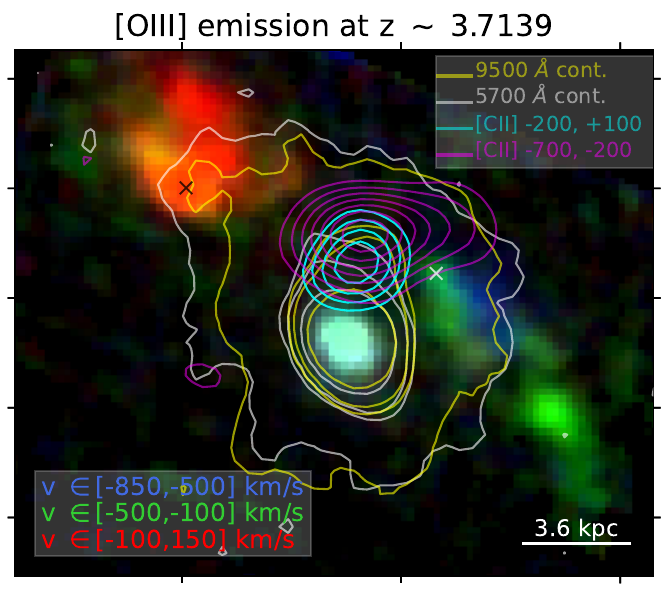}
\caption{\label{fig:elvelocity}{\bf Kinematics of the gas-rich group components Eastfield and Mr~West.} Left panel: Velocity histograms of the [OIII]$\lambda5007$ emission for Eastfield and Mr~West (top) and of the [CII] $\lambda$158~$\mu$m emission around Hyde (bottom). The zero-velocity corresponds to the redshift of Jekyll, $z=3.7135$.  Right panel: Three-color [OIII] maps obtained from the $R=2700$ 0.03 arcsec/pixel cube by integrating the ionized gas over specific velocity ranges (as labeled), in order to highlight the presence of multiple emitting regions in the surroundings of Jekyll and Hyde. These velocity ranges are shown as shaded regions in the left panel. The yellow and white contours display the stellar continuum emission as detected in the $R=100$ cube, in the intervals 4.00--4.95~$\mu$m and 2.5--3.0~$\mu$m, respectively. Levels in those contours are are 3, 30, 40, 80 $\sigma$. The cyan and magenta contours display the [CII] $\lambda$158~$\mu$m emission in two velocity ranges, as labeled and shown on the bottom-left panel. In the figure, we also mark with ‘x’ symbols the position of the two AGN candidates identified by \cite{2023arXiv231003067P}.}
\end{figure}

\begin{figure}
\centering
\includegraphics[trim=0.0cm 0.0cm 0.0cm 0.0cm,width=0.99\textwidth]{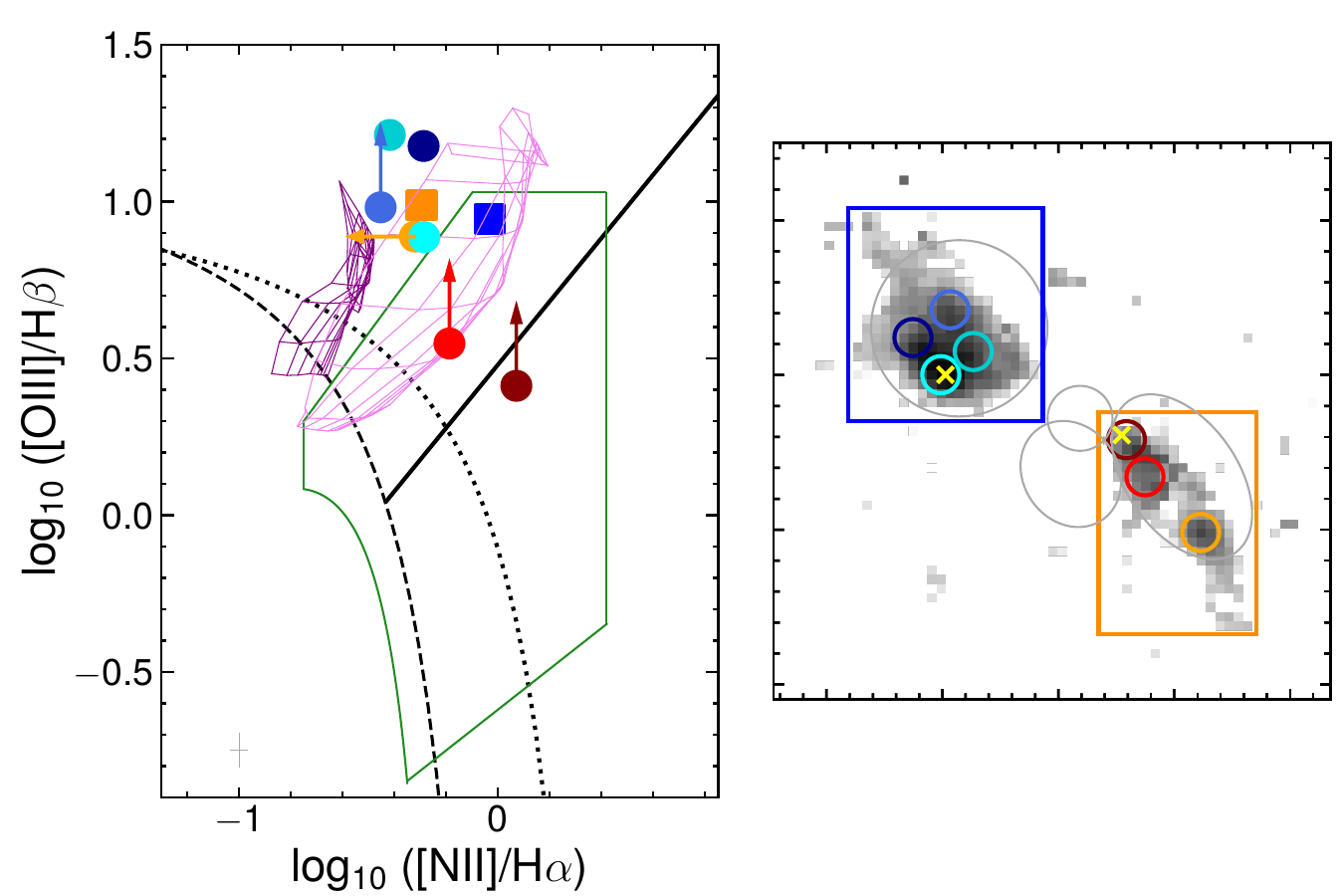}

\caption{{\bf Emission-line ratios in 2 dimensions of the gas-rich group components Eastfield and Mr~West.} Left panel: BPT diagnostic diagram of individual regions in Eastfield and Mr~West. The green solid line denotes the location in the diagram compatible with shock ionization from \cite{2016ApJS..224...38A}, while the purple and pink grids are sets of shock-plus-precursor models from \textsc{mappings v} \citep{2018ascl.soft07005S}, assuming $Z=0.01$ and 0.03, respectively, density n~=~1~cm$^{-3}$ and varying the shock velocity and strength of the magnetic field \cite{2018ascl.soft07005S}. Right panel: map of [OIII] emission highlighting the regions that are represented in the BPT diagram on the left. In the map, we also mark with ‘x’ symbols the position of the two AGN candidates identified by \cite{2023arXiv231003067P}.\label{fig:bpt}}
\end{figure}

\begin{figure}
\centering
\includegraphics[trim=3.5cm 0.0cm 0.0cm 0.0cm,width=1.0\textwidth]{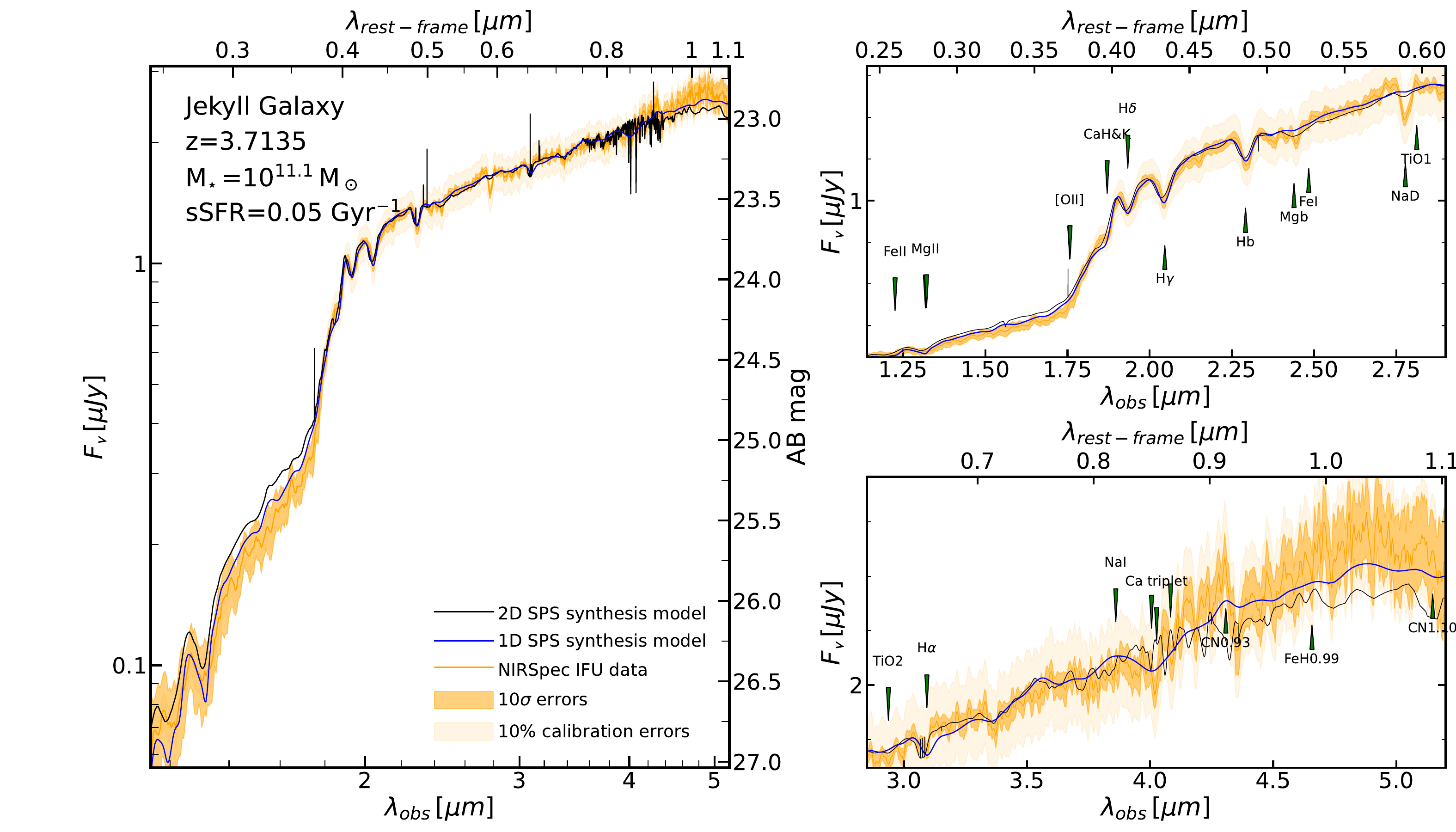}
\includegraphics[trim=3.5cm 0.0cm 0.0cm 0.0cm,width=1.0\textwidth]{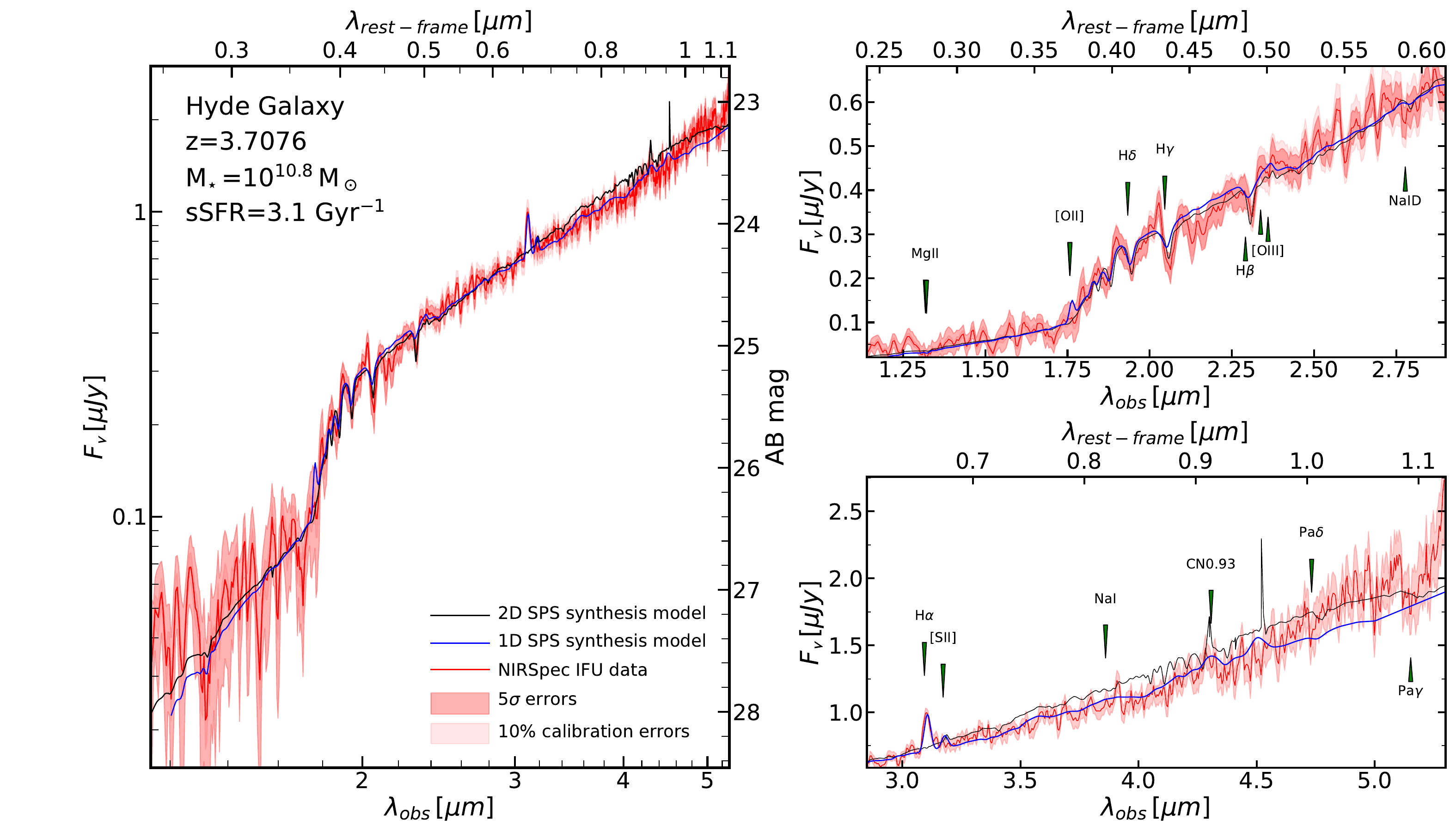}
\caption{\label{fig:jekyll_sed}{\bf Stellar population modeling from $R=100$ data in the Jekyll and Hyde galaxies.} Spectrum of the massive quiescent galaxy Jekyll (top panels) and Hyde (bottom panels) and fit to stellar population synthesis models. The figures show the sum of the fits to the individual spaxels of each galaxy, as shown in Figure~1. The SFH for these fits were shown in Figure~3. On the left, for each galaxy, we show the complete $R=100$ spectrum. The darker  shaded regions in orange and red show 10$\sigma$ errors, while the dimmer shaded region corresponds to 10\% of the flux, a conservative measurement of the absolute flux calibration of the NIRSpec IFU instrument. On the right, for each galaxy, we show 2 spectral ranges, marking with arrows some of the most interesting spectral features.}
\end{figure}

\begin{figure}
\centering
\includegraphics[trim=0.0cm 0.0cm 0.0cm 0.0cm,clip,width=0.9\textwidth]{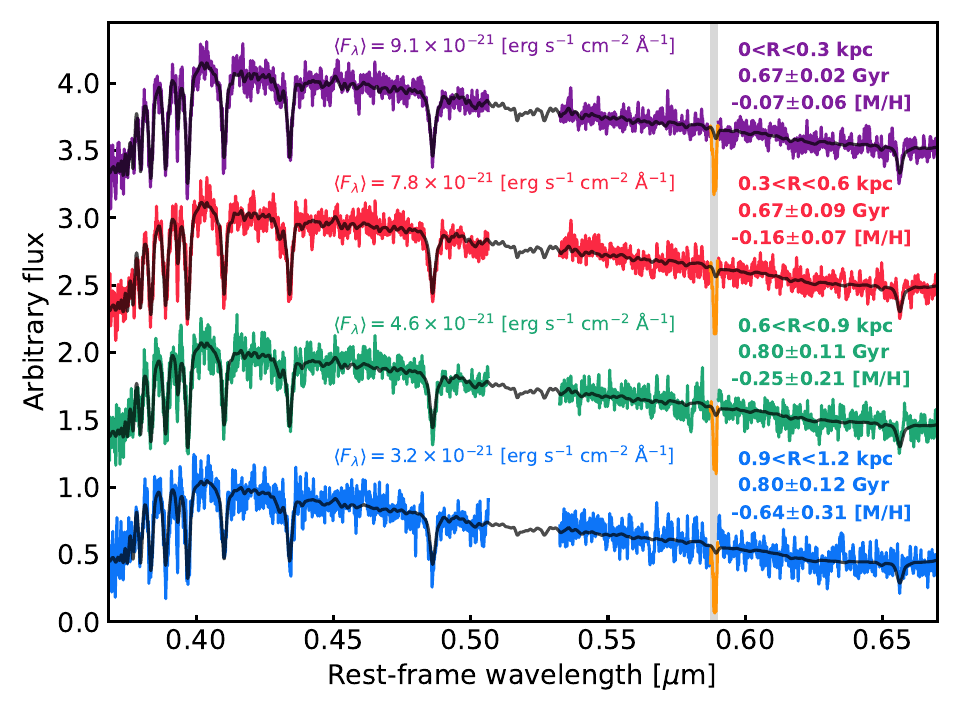}
\caption{\label{fig:metallicity}{\bf Spatially resolved stellar population modeling from $R=2700$ data of the Jekyll galaxy.} The figure shows the average spectra of Jekyll at four different radii. These $R=2700$ spectra were extracted using elliptical annuli, assuming a position angle PA=30$^o$  and an ellipticity of 0.3. The quoted distances are measured along the major axis. The average spectra at each radii (physical galactocentric radius given as a legend) was calculated by de-redshifting the spectra of individual spaxels out to the distance where radial velocities could be robustly measured, using the 2D information of the IFU to maximize the signal-to-noise of the stacked spectra. We provide the estimations for the stellar mass-weighted age and metallicity, as well as the average flux of each spectrum in the region of interest (rest-frame 3600-6700~\AA) for reference. The region around the Mgb feature (5170~\AA rest-frame, 2.43~$\mu$m observed) was masked during the fitting process because of the NIRSpec gap in that region, as well as the region of the NaD feature (rest-frame 5890~\AA) due to the strong blueshifted absorption indicating non-stellar origin (discussed in the main text as the Dr~Sodium component). Although the CaII K \& K lines are not fully reproduced either, we tested that masking them does not significantly affect the recovered ages and metallicities. The discrepancy between best-fitting models and data is related to the contribution of a small fraction of young stars not properly modeled.}
\end{figure}

\begin{figure}
\centering
\includegraphics[trim=0.5cm 0.0cm 1.cm 0.0cm,clip,width=0.5\textwidth]{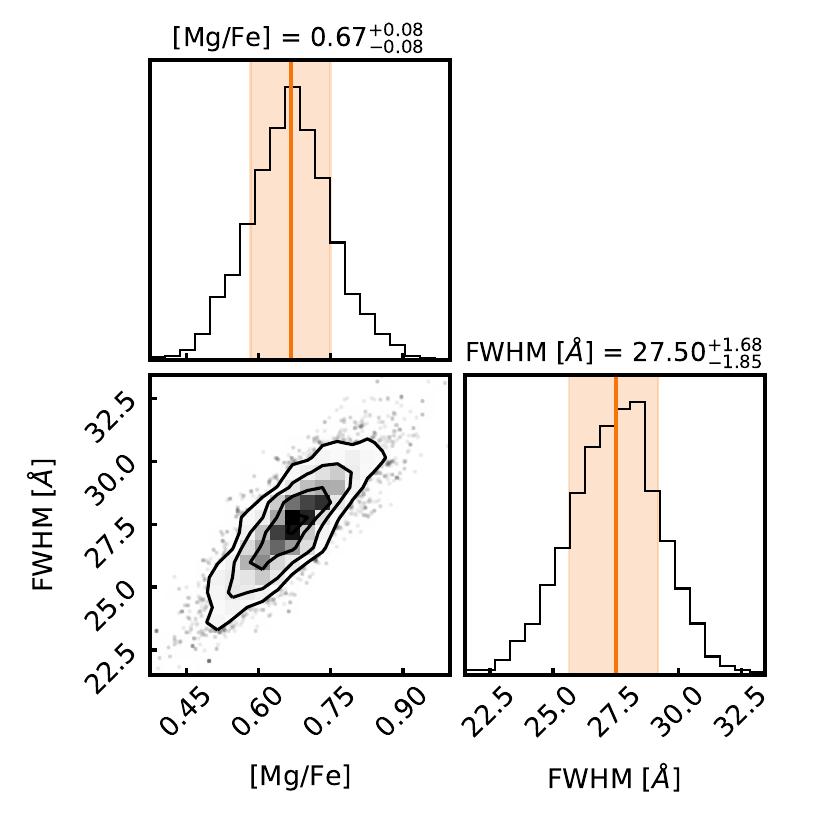}
\includegraphics[trim=0.0cm 0.0cm 0.0cm 0.0cm,clip,width=0.45\textwidth]{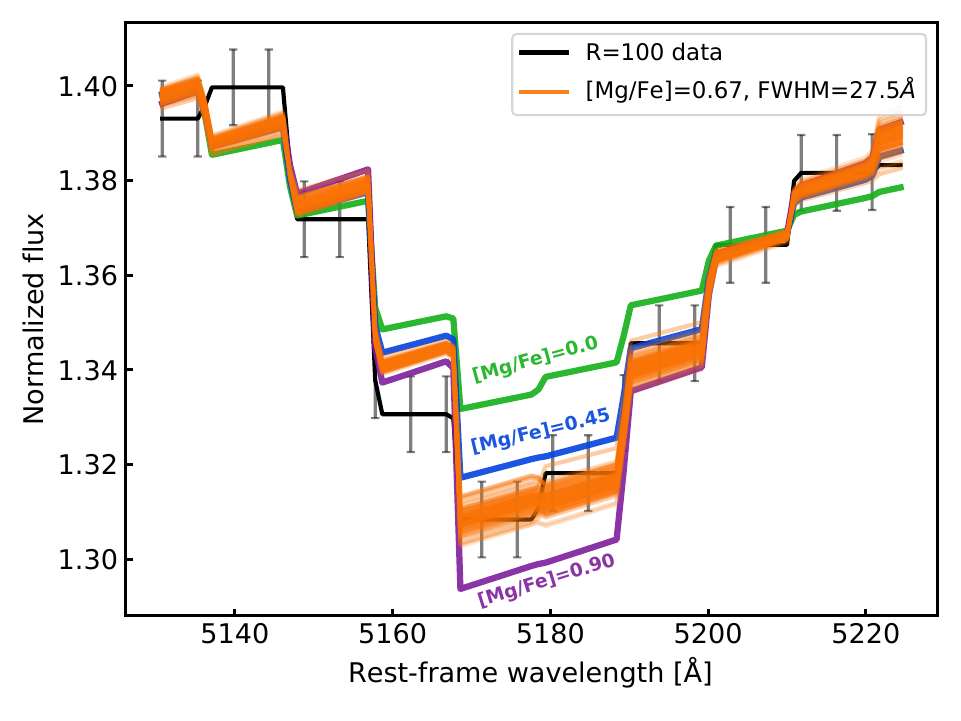}
\caption{\label{fig:mg-to-fe}{\bf Results of the Mg/Fe analysis of the $R=100$ spectrum of the Jekyll galaxy.} The dependence of the results with respect to the (rest-frame) spectral resolution are outlined, with the nominal value being 35~\AA. On the left, analysis of the posteriors of our fitting analysis. The contours show 1, 2, 3$\sigma$ on the posterior. On the right, we show the comparison of the data (in black) with models with $[Mg/Fe]=0.67\pm0.08$ (in orange) and other values of $[Mg/Fe]$ from 0.0 to 0.90 (for the same FWHM).}
\end{figure}

\begin{figure}
\centering
\includegraphics[trim=0.0cm 0.0cm 0.0cm 0.0cm,clip,width=1.0\textwidth]{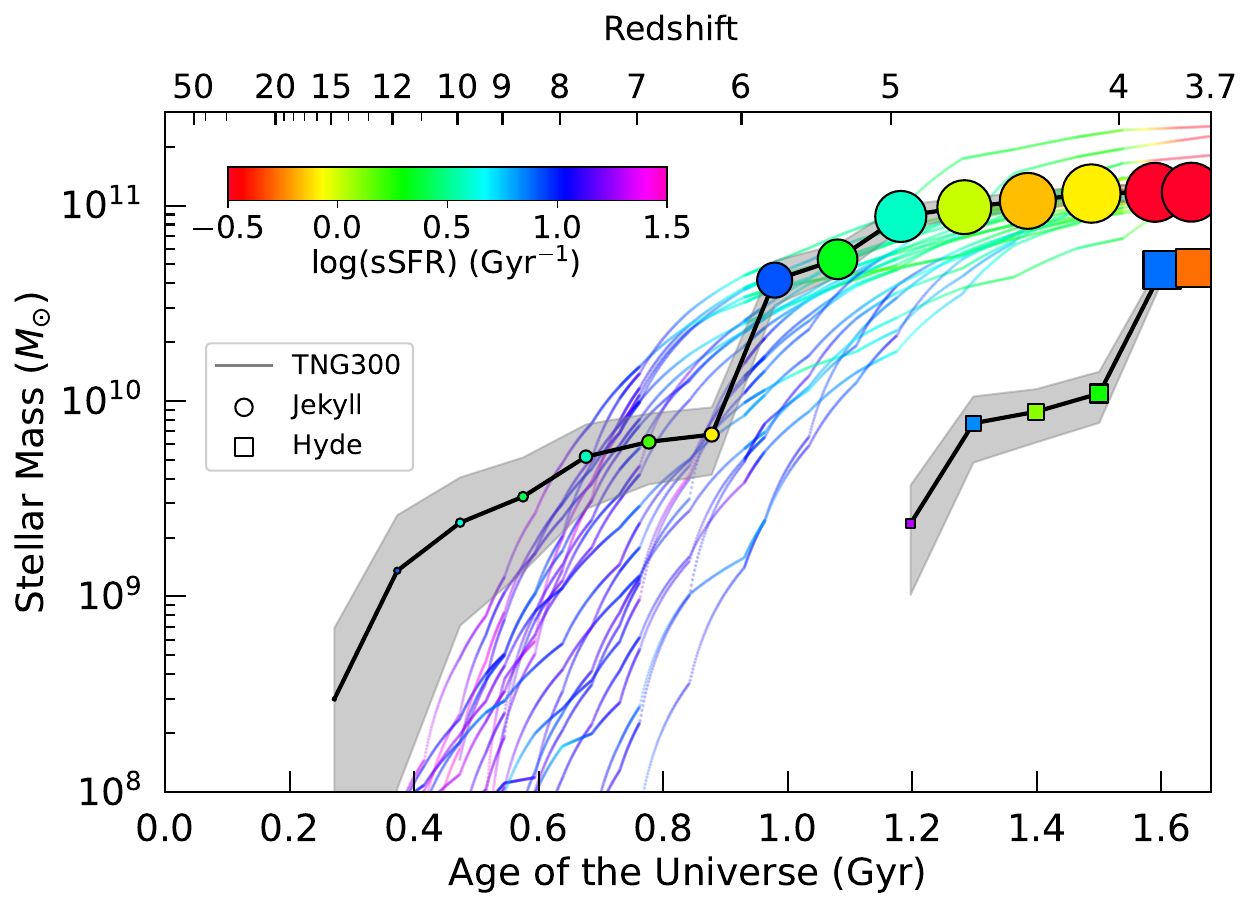}
\caption{\label{fig:tng-jekyll} {\bf Stellar mass evolution of the Jekyll, Hyde, and similar simulated galaxies.} Comparison of the mass accretion history of Jekyll (circles connected by a black solid line), Hyde (squares),
and massive quiescent galaxies in TNG300 (solid lines), based on the 2D analysis of the stellar populations which provide the SFHs shown in Figure~3.
The tracks are color-coded according to the  sSFR of each galaxy.
The symbol size is normalized to the total mass
of Jekyll at $z=3.7$. The results shown with data points for Jekyll and Hyde correspond to the most probable SFH shown in Figure~3, and the shaded regions refer to the standard deviation obtained with a Monte-Carlo approach.}
\end{figure}

\clearpage
\section*{Data availability}

The JWST NIRSpec data are available from the Mikulski Archive for Space Telescopes (MAST; \url{http://archive.stsci.edu}), under program IDs 1217 and 1837. The ALMA data is part of program 2015.A.00026.S, we obtained the calibrated measurement sets
from the EU ALMA Regional Centre (ARC, \url{https://www.eso.org/sci/facilities/alma/arc.html}). Any data used in this paper can be provided upon request to the first author.

\section*{Code availability}

JWST NIRSpec data were reduced using the JWST Pipeline (Version 1.8.2,
reference mapping 1105; \url{https://github.com/spacetelescope/jwst}). NIRSpec
data inspection used the Mosviz visualization tool (\url{https://jdaviz.readthedocs.io/en/latest/mosviz/index.html}). We used the ALMA data reduction software
CASA v6.2.1 \cite{CASA}.  This research made use of Astropy, a community-developed core Python package for Astronomy (\url{https://www.astropy.org/}), and {\sc statmorph} (\url{https://statmorph.readthedocs.io/en/latest/}). Three different codes were used for stellar population synthesis modeling of galaxy spectro-photometric data: Synthesizer
\cite{2003MNRAS.338..508P,2008ApJ...675..234P}; PPXF \cite{2004PASP..116..138C}; and prospector \cite{johnson21,leja19}.

\section*{Acknowledgments}
We thank the referees for very useful comments that improved the interpretation of the data. We want to thank Karl Glazebrook, who discovered  the Jekyll galaxy, for his input on the nature of the system that improved this paper, as well as Elena Bertola for their useful comments on the manuscript. We thank G. Mazzolari for helping with the analysis of the radio data. We acknowledge support from grants PID2022-139567NB-I00, PID2021-127718NB-I00, and PID2022-140483NB-C22  funded by Spanish Ministerio de Ciencia e Innovaci\'on MCIN/AEI/10.13039/501100011033,
FEDER {\it Una manera de hacer
Europa}, the Recovery, Transformation and Resilience Plan from the Spanish State, and by NextGenerationEU from the European Union through the Recovery and Resilience Facility, as well as the Programa Atracci\'on de Talento de la Comunidad de Madrid (Spain) via grant 2018-T2/TIC-11715.
We are grateful for the support of ALMA EU archive. The project leading to this publication has received support from ORP, that is funded by the European Union’s Horizon 2020 research and innovation programme under grant agreement No 101004719 [ORP].
FDE, RM and JS acknowledge support by the Science and Technology Facilities Council (STFC), by the ERC through Advanced Grant 695671 ``QUENCH'', and by the UKRI Frontier Research grant RISEandFALL.
FDE is grateful to C. Conroy for sharing the high-resolution C3K/MIST library. RM also acknowledges funding from a research professorship from the Royal Society.
H{\"U} acknowledges funding by the European Union (ERC APEX, 101164796). Views and opinions expressed are however those of the authors only and do not necessarily reflect those of the European Union or the European Research Council Executive Agency. Neither the European Union nor the granting authority can be held responsible for them.
AJB acknowledges funding from the ``FirstGalaxies'' Advanced Grant from the European Research Council (ERC) under the European Union's Horizon 2020 research and innovation programme (Grant agreement No. 789056).
SC, EP and GV acknowledge support by European Union’s HE ERC Starting Grant No. 101040227 - WINGS.
GC acknowledges the support of the INAF Large Grant 2022 ``The metal circle: a new sharp view of the baryon
cycle up to Cosmic Dawn with the latest generation IFU facilities''. I.L. acknowledges support from PRIN-MUR project “PROMETEUS”  financed by the European Union -  Next Generation EU, Mission 4 Component 1 CUP B53D23004750006.
%
%


\section*{Author contribution}

PGP-G and FD'E led the analysis and interpretation effort, writing the core of the manuscript. BRdP carried out the kinematic measurements analyzing the nebular emission. MP and H\"U carried out the JWST/NIRSpec data calibration and helped with active galactic nuclei diagnostic plots and properties. RM, SA, and GC contributed with the overall interpretation of the properties of the galaxy group. IL compiled and analyzed the ALMA data. GB analyzed the stellar populations in the system. LC searched for analogs of the galaxies in this system in galaxy formation simulations. IM-N analyzed the high spectral resolution data to infer detailed ages, metallicities, and abundances of the quiescent galaxy. JD and DM contributed with the reduction and analysis of the JWST/NIRCam imaging data. AJB, SC, SC, CJW, T.B, EP, JS, and GV, and all others authors, contributed to manuscript writing, discussions about the results, as well as the development of the NIRSpec instrument and the planning of the guaranteed time observations during more than two decades.


\section*{Competing Interests Statement}

The authors declare no competing interests.





\end{document}